\documentclass[pra,aps,twocolumn,superscriptaddress,nofootinbib]{revtex4}
\usepackage{hyperref}
\usepackage{bm}
\usepackage{epsfig}
\usepackage{amssymb}
\usepackage{graphicx}
\usepackage{mathrsfs}
\usepackage{amsmath}
\usepackage{epsfig}

\begin{document}
\title{Gaussian intrinsic entanglement}

\author{Ladislav Mi\v{s}ta, Jr.}
\affiliation{Department of Optics, Palack\' y University,
17.~listopadu 12,  771~46 Olomouc, Czech Republic}

\author{Richard Tatham}
\affiliation{Department of Optics, Palack\' y University,
17.~listopadu 12,  771~46 Olomouc, Czech Republic}

\begin{abstract}

We introduce a cryptographically motivated quantifier of
entanglement in bipartite Gaussian systems called {\it Gaussian
intrinsic entanglement} (GIE). The GIE is defined as the optimized
mutual information of a Gaussian distribution of outcomes of
measurements on parts of a system, conditioned on the outcomes of
a measurement on a purifying subsystem. We show that GIE vanishes
only on separable states and exhibits monotonicity under Gaussian
local trace-preserving operations and classical communication. In
the two-mode case we compute GIE for all pure states as well as
for several important classes of symmetric and asymmetric mixed
states. Surprisingly, in all of these cases, GIE is equal to
Gaussian R\'{e}nyi-2 entanglement. As GIE is operationally
associated to the secret-key agreement protocol and can be
computed for several important classes of states it offers a
compromise between computable and physically meaningful
entanglement quantifiers.

\end{abstract}

\maketitle

Since its discovery, entanglement has transitioned from a mere
paradoxical feature of quantum mechanics \cite{Einstein_35} to a
powerful resource for communication and computing. The natural quest
to discover all facets of entanglement revealed the need not only to
verify its presence but also to quantify it. Some crucial
properties of entanglement such as monogamy \cite{Coffman_00} are
quantitative and therefore cannot be described without introducing
entanglement measures. Further, entanglement measures are required
for the characterization of entangling gates \cite{O'Brien_04} and
set the bounds one has to surpass in experiments into some key
quantum information protocols such as entanglement distillation
\cite{Yamamoto_03}.

Existing entanglement measures either possess a good
operational meaning or are computable but not both. The first
kind of measure is best exemplified by the distillable
entanglement \cite{Bennett_96}, which quantifies the pure-state
entanglement one can distill from a shared quantum state but is
difficult to compute. At the opposite extreme is the logarithmic
negativity \cite{Vidal_02,Eisert_PhD}, which is computable for any
state but lacks a coherent operational interpretation. However,
to assess the utility of a given entangled state in practical tasks, we
need to develop entanglement measures which are both computable
and physically meaningful. Unfortunately, except for the entanglement of formation
\cite{Bennett_96}, which quantifies how much pure-state
entanglement one needs to create a shared quantum state and which
can be computed for two qubits \cite{Wootters_98} and Gaussian
states \cite{Giedke_03a,Marian_08}, no such measure is currently
known.

One way to probe the gap is to quantify entanglement in the
context of classical secret key agreement \cite{Maurer_93}. In
this cryptographic protocol, three random variables $A,B$ and $E$
distributed according to $P(A,B,E)$ are held by two honest
parties, Alice and Bob, and an adversary, Eve. Alice and Bob are
connected by a public communication channel and their goal is to
generate a secret key, that is a common string of random bits
about which Eve has practically no information. For the key
agreement to be possible it is necessary that Alice and Bob share
correlations which cannot be distributed by public communication,
i.e., secret correlations. A useful quantifier of secret
correlations is the so called intrinsic information
\cite{Maurer_99}
\begin{eqnarray}\label{xI}
I(A; B\downarrow E)=\mathop{\mbox{inf}}_{E\rightarrow
\tilde{E}}[I(A; B| \tilde{E})],
\end{eqnarray}
where $I(A; B|\tilde{E})$ is the conditional mutual information
and the infimum is taken over all channels
$E\rightarrow\tilde{E}$. The intrinsic information is an upper
bound (not always tight \cite{Renner_03}) on the rate at which
a secret key can be generated from the
investigated distribution and what is more, it is conjectured that
it is equal to a secret key rate in the modified key agreement
protocol called public Eve scenario \cite{Christandl_04,PES}.

The intrinsic information (\ref{xI}) can be used to quantify
entanglement in a quantum state $\rho_{AB}$. This is accomplished
by projective measurements in some bases $\{|A\rangle\}$ and
$\{|B\rangle\}$, and a generalized measurement with a generating
set $\{|E\rangle\}$ on subsystems $A,B$ and $E$ of a purification
$|\Psi\rangle$ of the state,
$\mbox{Tr}_{E}|\Psi\rangle\langle\Psi|=\rho_{AB}$. If the state
$\rho_{AB}$ is entangled (separable) and the basis (set)
$\{|A\rangle\}$ and $\{|B\rangle\}$ ($\{|E\rangle\}$) is chosen
suitably, the obtained distribution $P(A,B,E)=|\langle A|\langle
B|\langle E|\Psi\rangle|^{2}$ has strictly positive (zero)
intrinsic information for any choice of the set (basis)
$\{|E\rangle\}$ ($\{|A\rangle\}$ and $\{|B\rangle\}$)
\cite{Gisin_00}. Thus, to faithfully map entanglement onto secret
correlations the optimized intrinsic information \cite{Gisin_00},
$\mu(\rho_{AB})=\mathop{\mbox{inf}}_{\left\{|E\rangle,|\Psi\rangle\right\}}
\{\mathop{\mbox{sup}}_{\left\{|A\rangle,|B\rangle\right\}}
\left[I\left(A;B\downarrow E\right)\right]\}$, has to be taken,
which exhibits some properties of an entanglement measure, namely
equality to the von Neumann entropy on pure states and convexity.
It might seem that $\mu$ is a good candidate for the sought
entanglement measure but it has two drawbacks. First, it is not
known whether it is non-increasing under local operations and
classical communication (LOCC) as it should \cite{Vidal_00}.
Second, so far it has been computed only for a two-qubit Werner
state \cite{Gisin_00}.

In this Letter we propose a quantifier of bipartite entanglement
called {\it intrinsic entanglement} (IE) defined as
\begin{equation}\label{xEdownarrow}
E_{\downarrow}(\rho_{AB})=\mathop{\mbox{sup}}_{\left\{|A\rangle,|B\rangle\right\}}
\left\{\mathop{\mbox{inf}}_{\left\{|E\rangle,|\Psi\rangle\right\}}\left[I\left(A;
B\downarrow E\right)\right]\right\}.
\end{equation}
The IE contains a reverse order of optimization in comparison with
the quantifier $\mu$ and hence $E_{\downarrow}\leq\mu$ due to the
max-min inequality \cite{Boyd_04}. The main advantage of IE is
that one can compute it easier than $\mu$ as we show below. We
focus on IE for an important class of Gaussian states. These
states are the backbone of quantum information technologies based
on continuous variables \cite{Weedbrook_12} and occur as ground or
thermal state of any bosonic quantum system in a ``linearized''
approximation \cite{Schuch_06}. Gaussian states can be also easily
prepared, manipulated and measured in many experimental platforms
encompassing light, atomic ensembles, trapped ions or
optomechanical systems \cite{Cerf_07}. Unfortunately, evaluation
of IE for Gaussian states involves complex optimization over
generally non-Gaussian measurements and states, and thus some
simplifications are needed. First, we restrict Alice and Bob to
Gaussian measurements. We do that because a scheme generating
distribution $P$ represents the first stage of a quantum key distribution protocol
(with an individual attack) and in protocols based on Gaussian
states honest parties typically perform Gaussian measurements
\cite{Weedbrook_12}. Second, we assume that it is optimal to use a
Gaussian measurement and channel on Eve's side. This assumption is
plausible as Gaussian attacks are optimal \cite{Grosshans_04} with
respect to the lower bound on the secret key rate for all
important Gaussian protocols.

Here we thus investigate the so called Gaussian IE (GIE) defined
by Eqs.~(\ref{xI}) and (\ref{xEdownarrow}), where all states,
measurements and channels are Gaussian. We show that for a
Gaussian state $\rho_{AB}$, the GIE is equal to the optimized
mutual information of a distribution of outcomes of Gaussian
measurements on subsystems $A$ and $B$ of a conditional state
obtained by a Gaussian measurement on subsystem $E$ of a Gaussian
purification of the state $\rho_{AB}$. Next, we prove that GIE is
faithful, i.e., it vanishes iff $\rho_{AB}$ is separable, and it
does not increase under Gaussian local trace-preserving operations
and classical communication (GLTPOCC). Finally, we compute GIE
analytically for several important classes of mixed two-mode
symmetric and asymmetric Gaussian states. As the optimum in GIE is
always reached by feasible homodyne and heterodyne detection, it
is an experimentally meaningful quantity which stays in line with
other optimized quantities such as Gaussian quantum discord
\cite{Giorda_10,Adesso_10,Pirandola_14}, where the optimum is also
often attained by homodyning or heterodyning. Remarkably, we find
further, that the calculated GIE is always equal to an important
measure of Gaussian entanglement called Gaussian R\'{e}nyi-2 (GR2)
entanglement \cite{Adesso_12}, which is defined as a convex roof
of the pure-state R\'{e}nyi-2 entropy of entanglement. The GR2
entanglement is a proper and natural measure of Gaussian
entanglement being monotonic under all Gaussian LOCC (GLOCC) and
monogamous. Additionally, the GR2 entanglement is additive on
two-mode symmetric states and finds interpretation in terms of a
phase-space sampling entropy for Wigner function
\cite{Adesso_12,Buzek_95}. Our findings lead us to a conjecture,
that GIE and GR2 entanglement are equal on all Gaussian states. If
the conjecture is true, all properties of the latter quantity
extend to the former and vice versa, thereby providing us with an
exceptional measure of Gaussian entanglement which has
cryptographic interpretation and many important properties, and
which can be computed in many cases.

We consider quantum systems with infinite-dimensional Hilbert
state space, e.g., light modes. An $n$-mode system is
characterized by a vector of quadratures
$\xi=(x_1,p_1,\ldots,x_n,p_n)^{T}$ whose components obey the
canonical commutation rules $[\xi_j,\xi_k]=i(\Omega_n)_{jk}$ with
$\Omega_{n}=\oplus_{j=1}^{n}i\sigma_{y}$, where $\sigma_{y}$ is
the Pauli-$y$ matrix. Gaussian states are fully described by a
covariance matrix (CM) $\gamma$ with entries
$\gamma_{jk}=\langle\xi_{j}\xi_k+\xi_k\xi_j\rangle-2\langle\xi_j\rangle\langle\xi_k\rangle$
and by a vector of first moments $\langle\xi\rangle$, which is
irrelevant in the present entanglement analysis and so is assumed
to be zero. We use Gaussian unitary operations which are for $n$
modes represented at the level of CMs by a real $2n\times 2n$
symplectic matrix $S$ fulfilling $S\Omega_{n}S^{T}=\Omega_{n}$.
We restrict ourselves to Gaussian measurements described by the
positive operator valued measure \cite{Fiurasek_07}
\begin{equation}\label{xPOVMn}
\Pi(d)=(2\pi)^{-n}D(d)\Pi_0 D^\dagger(d),
\end{equation}
which satisfies the completeness condition
$\int_{\mathbb{R}_{2n}}\Pi(d){\rm d}^{2n}d=\openone$, where ${\rm
d}^{2n}d=\Pi_{l=1}^{n}{\rm d}d_{l}^{(x)}{\rm d}d_{l}^{(p)}$. Here,
the seed element $\Pi_0$ is a normalized density matrix of a
generally mixed $n$-mode Gaussian state with zero first moments
and CM $\Gamma$, $D(d)=\exp(-id^{T}\Omega_{n}\xi)$ is the
displacement operator, and
$d=(d_{1}^{(x)},d_{1}^{(p)},\ldots,d_{n}^{(x)},d_{n}^{(p)})^{T}\in\mathbb{R}_{2n}$
is a vector of measurement outcomes.

{\it Simplification of GIE}.---Initially we show that the
assumption of Gaussianity of all states, measurements, and the
channel $E\rightarrow\tilde{E}$, considerably simplifies the
quantity (\ref{xEdownarrow}). Assume that
$\rho_{AB}\equiv\rho_{A_1\ldots A_{N}B_{1}\ldots B_{M}}$ of
Eq.(\ref{xEdownarrow}) is an $(N+M)$-mode Gaussian state with CM
$\gamma_{AB}$. Let us further assume that $|\Psi\rangle_{ABE}$ is
a Gaussian purification of the state with CM $\gamma_{\pi}$, which
contains $K$ purifying modes $E_1,E_2,\ldots,E_K$. Consider now,
that the subsystems $A,B$ and $E$ are distributed among Alice, Bob
and Eve, who carry out local Gaussian measurements (\ref{xPOVMn})
characterized by covariance matrices (CMs) $\Gamma_{A},\Gamma_{B}$
and $\Gamma_{E}$, respectively. As a result, the participants
share a zero-mean Gaussian distribution $P(d_A,d_B,d_E)$ of
measurement outcomes $d_A,d_B$ and $d_E$ with a classical
covariance matrix (CCM) \cite{CCM} $\Sigma$ expressed with respect
to the $AB|E$ partitioning as
\begin{eqnarray}\label{xSigma}
\Sigma=\left(\begin{array}{cc}
\gamma_{AB}+\Gamma_{A}\oplus\Gamma_{B} & \gamma_{ABE}\\
\gamma_{ABE}^{T} & \gamma_{E}+\Gamma_{E}\\
\end{array}\right)\equiv\left(\begin{array}{cc}
\alpha & \beta\\
\beta^{T} & \delta\\
\end{array}\right),
\end{eqnarray}
where $\gamma_{AB}, \gamma_{ABE}$ and $\gamma_{E}$ are blocks of
the CM $\gamma_{\pi}$ according to the same partitioning. In what
follows, we analyze GIE defined by Eq.~(\ref{xEdownarrow}), where
the role of the distribution $P(A,B,E)$ is played by the Gaussian
distribution $P(d_A,d_B,d_E)$ and the optimization is performed
over Gaussian channels $E\rightarrow\tilde{E}$ and CMs
$\gamma_{\pi}$ and $\Gamma_{A,B,E}$.

First, we identify the conditional mutual information $I\left(A;
B| E\right)$ of Eq.~(\ref{xI}) for the distribution
$P(d_A,d_B,d_E)$. According to definition \cite{Cover_06}
$I(A;B|E)$ is the standard mutual information $I_{c}(A;B)$ of the
conditional distribution $P(d_A,d_B|d_E)$ averaged over the
distribution of the variable $d_E$. The distribution is Gaussian
with a CCM given by the Schur complement \cite{Horn_85,Giedke_02}
of CCM (\ref{xSigma})
\begin{eqnarray}\label{xsigmaAB}
\sigma_{AB}&=&\gamma_{AB}+\Gamma_{A}\oplus\Gamma_{B}-\gamma_{ABE}\frac{1}{\gamma_{E}+\Gamma_{E}}\gamma_{ABE}^{T},
\end{eqnarray}
where the inverse is to be understood generally as the
pseudoinverse.
Making use of the formula for mutual information of a bivariate
Gaussian distribution \cite{Gelfand_57}, we arrive at
$I_{c}(A;B)=f(\gamma_{\pi},\Gamma_{A},\Gamma_{B},\Gamma_{E})$,
where
\begin{equation}\label{xf}
f(\gamma_{\pi},\Gamma_{A},\Gamma_{B},\Gamma_{E})=\frac{1}{2}\ln\left(\frac{\mbox{det}\sigma_{A}\mbox{det}\sigma_{B}}
{\mbox{det}\sigma_{AB}}\right)
\end{equation}
with $\sigma_{A,B}$ being local submatrices of CCM
(\ref{xsigmaAB}). From Eq.~(\ref{xf}) it then follows that
$I_{\rm c}(A;B)$ is independent of $d_E$ and
hence $I(A;B|E)=I_{\rm
c}(A;B)=f(\gamma_{\pi},\Gamma_{A},\Gamma_{B},\Gamma_{E})$.

We next prove that the channel $E\rightarrow\tilde{E}$ in
Eq.~(\ref{xI}) can be integrated into Eve's measurement. Again, we
assume a Gaussian channel \cite{Caruso_08},
$\tilde{d}_{E}=Xd_{E}+y$, mapping the $2K\times 1$ vector $d_{E}$
of Eve's measurement outcomes onto a new $L\times 1$ vector
$\tilde{d}_{E}$. Here $X$ is a real matrix and
$y=(y_1,y_2,\ldots,y_L)^{T}$ is a random vector obeying a zero
mean Gaussian distribution with CCM $Y$ with elements
$Y_{ij}=2\langle y_{i}y_{j}\rangle$. The channel transforms the
CCM (\ref{xsigmaAB}) to
\begin{eqnarray}\label{xtildesigmaAB}
\tilde{\sigma}_{AB}&=&\alpha-\beta X^{T}\frac{1}{X\delta
X^{T}+Y}X\beta^{T},
\end{eqnarray}
where $\alpha,\beta$ and $\delta$ are blocks of CCM (\ref{xSigma}).
With the help of the singular value decomposition \cite{Horn_85}
of matrix $X$, CCM (\ref{xtildesigmaAB}) can be recast after some
algebra into the form (\ref{xsigmaAB}) with CM $\Gamma_{E}$
replaced with a different CM (see \cite{Mista_15} for the explicit
form of the new CM). Therefore, a Gaussian measurement on Eve's
system followed by a Gaussian channel on outcomes of the
measurement is equivalent to another Gaussian measurement, which
concludes the proof.

Further simplification follows from the invariance of CCM
(\ref{xsigmaAB}) under a change of which purification state is
used, accompanied by a corresponding change to Eve's measurement.
Namely, for any two CMs $\gamma_{\pi}$ and $\bar{\gamma}_{\pi}$ of
purifications with $K$-mode and $\bar{K}$-mode purifying subsystem
$E$ and $\bar{E}$, where $K\leq\bar{K}$, there is a symplectic
matrix on subsystem $\bar{E}$ which brings CM $\bar{\gamma}_{\pi}$
to CM $\gamma_{\pi}\oplus[\openone^{\oplus(\bar{K}-K)}]$
\cite{Mista_15,Giedke_03b,Magnin_10}. As a result, for CM
$\bar{\gamma}_{\pi}$ and CM $\bar{\Gamma}_{\bar{E}}$ of a
measurement on subsystem $\bar{E}$, which possess CCM
$\bar{\sigma}_{AB}$, Eq.~(\ref{xsigmaAB}), there is for CM
$\gamma_{\pi}$ a CM $\Gamma_{E}$ of a measurement on subsystem $E$
giving $\sigma_{AB}=\bar{\sigma}_{AB}$, and vice versa
\cite{Mista_15}. Accordingly, there is a free choice over which
Gaussian purification of $\rho_{AB}$ to work with.

The proposed quantity GIE is defined as the conditional mutual
information $I(A;B|\tilde{E})$ given in Eq.~(\ref{xf}), where
$\sigma_{AB}$ is replaced with $\tilde{\sigma}_{AB}$,
Eq.~(\ref{xtildesigmaAB}), which is first minimized with respect to
all Gaussian channels $E\rightarrow\tilde{E}$ and subsequently CMs
$\Gamma_{E}$ and $\gamma_{\pi}$ of measurements and purifications,
respectively, and then maximized with respect to all pure-state
CMs $\Gamma_{A}$ and $\Gamma_{B}$. As any Gaussian channel can be
incorporated into Eve's measurement, we can omit the minimization
with respect to the channels without loss of generality.
Furthermore, as for any purification and measurement on subsystem
$E$ there is a measurement on subsystem $E$ of a fixed
purification giving the same conditional mutual information
(\ref{xf}), we can further restrict ourselves in the definition of
GIE to a fixed purification and minimization with respect to all
CMs $\Gamma_{E}$. Consequently, GIE simplifies to
\begin{equation}\label{xEGdownarrow}
E_{\downarrow}^{G}\left(\rho_{AB}\right)=\mathop{\mbox{sup}}_{\Gamma_{A},\Gamma_{B}}
\mathop{\mbox{inf}}_{\Gamma_{E}}f(\gamma_{\pi},\Gamma_{A},\Gamma_{B},\Gamma_{E}),
\end{equation}
where $f$ is given in Eq.~(\ref{xf}), $\gamma_{\pi}$ is CM of a
fixed purification and the infimum (supremum) is taken over all
CMs $\Gamma_{E}$ ($\Gamma_{A,B}$) of measurements on subsystem $E$
($A$ and $B$).


{\it Faithfulness}.---We first prove that GIE vanishes iff
$\rho_{AB}$ is separable. The proof closely follows a similar
proof for intrinsic information given in Ref.~\cite{Gisin_00}. The
``only if'' part has been proved in Ref.~\cite{Mista_15}. It
follows from the fact, that any separable Gaussian state has a
Gaussian purification which can be projected onto a product state
of subsystems $A$ and $B$ by a suitable measurement on subsystem
$E$. Hence, after the measurement the CCM (\ref{xsigmaAB}) reduces
to
$\sigma_{AB}=(\gamma_{A}+\Gamma_{A})\oplus(\gamma_{B}+\Gamma_{B})$
for any local Gaussian measurements on subsystems $A$ and $B$,
where $\gamma_{A,B}$ denote local CMs of the product state. This
implies that the mutual information (\ref{xf}) vanishes for any CMs
$\Gamma_{A}$ and $\Gamma_{B}$, and therefore
$E_{\downarrow}^{G}\left(\rho_{AB}\right)=0$ as required.

The ``if'' part can be proved by contradiction. Let
$E_{\downarrow}^{G}\left(\rho_{AB}\right)=0$ for some entangled
state $\rho_{AB}$. Then, for any CMs $\Gamma_{A}$ and $\Gamma_{B}$
there is a CM $\Gamma_{E}$ such that the mutual information in
Eq.~(\ref{xf}) vanishes. This is equivalent to the statistical
independence of the variables obeying the respective bivariate
Gaussian distribution with CCM $\sigma_{AB}$ \cite{Cover_06},
which implies that $\sigma_{AB}=\sigma_{A}\oplus\sigma_{B}$.
Hence, for a Gaussian measurement $\Pi_{E}(d_{E})$ with CM
$\Gamma_{E}$ the corresponding (unnormalized) conditional state
$\mbox{Tr}_{E}[|\Psi\rangle\langle\Psi|\Pi_{E}(d_{E})]$
factorizes. By integrating the latter state over all measurement
outcomes $d_E$ and taking into account the completeness condition
for the measurement $\Pi_{E}(d_E)$ one gets an expression of the
state $\rho_{AB}$ in the form of a convex mixture of product
states and therefore the state is separable, which is a
contradiction. Thus, equality
$E_{\downarrow}^{G}\left(\rho_{AB}\right)=0$ implies separability
of $\rho_{AB}$ which accomplishes the proof.

\begin{figure}[tbh]
\includegraphics[width=1\columnwidth]{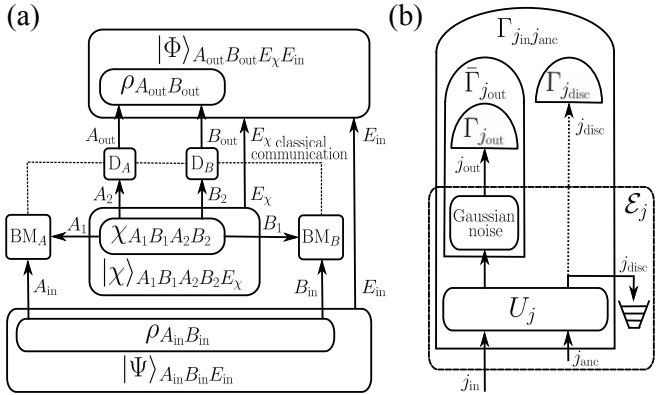}
\caption{(a) Construction of the purification of the output state
$\rho_{A_{\rm out}B_{\rm out}}$ via teleportation.
$\mbox{BM}_{j}$: Bell measurement on subsystem $(j_{\rm
in}j_{1})$, $\mbox{D}_{j}$: displacement of subsystem $j_{2}$. (b)
Decomposition of operation $\mathcal{E}_{j}$ and its integration
into a measurement.}\label{fig_1}
\end{figure}

{\it Monotonicity}.--- For GIE to be a good Gaussian entanglement
measure it should not increase under GLOCC
\cite{Vidal_00,Wolf_04}. This means, that if such an operation
$\mathcal{E}$ maps an input Gaussian state $\rho_{A_{\rm in}B_{\rm
in}}$ onto an output Gaussian state $\rho_{A_{\rm out}B_{\rm
out}}$, then
\begin{equation}\label{xmonotonicity}
E_{\downarrow}^{G}\left(\rho_{A_{\rm in}B_{\rm in}}\right)\geq
E_{\downarrow}^{G}\left(\rho_{A_{\rm out}B_{\rm out}}\right).
\end{equation}
We here outline the proof of inequality (\ref{xmonotonicity}) for
the subset of GLOCC given by GLTPOCC (see \cite{Mista_15} for the
detailed proof). First, we construct a suitable purification of
the output state $\rho_{A_{\rm out}B_{\rm out}}$. For this
purpose, we use realization of the operation $\cal{E}$ by a
continuous-variable teleportation protocol \cite{Braunstein_98},
where a quantum channel is a Gaussian state $\chi$ representing
the operation \cite{Giedke_02,Jamiolkowski_72,Fiurasek_02} (see
Fig.~\ref{fig_1}(a)). Here, the input state $\rho_{A_{\rm
in}B_{\rm in}}$ is teleported via input subsystems $A_{1}$ and
$B_{1}$ of the state $\chi_{A_{1}B_{1}A_{2}B_{2}}$ to the output
subsystems $A_{2}$ and $B_{\rm 2}$ by Bell measurements on
composite subsystems $(A_{\rm in}A_{1})$ and $(B_{\rm in}B_{1})$.
The measurements comprise separate measurements of the difference of
the $x$-quadratures and the sum of the $p$-quadratures on each
corresponding pair of modes. After the measurements and suitable
displacements of subsystems $A_{2}$ and $B_{2}$, the output state
$\rho_{A_{\rm out}B_{\rm out}}$ is obtained on subsystems $A_{\rm
out}$ and $B_{\rm out}$. Now, by replacing states $\rho_{A_{\rm
in}B_{\rm in}}$ and $\chi_{A_{1}B_{1}A_{2}B_{2}}$ with their
purifications $|\Psi\rangle_{A_{\rm in}B_{\rm in}E_{\rm in}}$ and
$|\chi\rangle_{A_{1}B_{1}A_{2}B_{2}E_{\chi}}$, respectively, and
teleporting the respective parts of the purifications, we get the
sought purification $|\Phi\rangle_{A_{\rm out}B_{\rm
out}E_{\chi}E_{\rm in}}$ of the output state $\rho_{A_{\rm
out}B_{\rm out}}$ (see Fig.~\ref{fig_1}(a)).
As the operation $\mathcal{E}$ is GLTPOCC, the state $\chi$ is a
Gaussian mixture of displaced Gaussian product states
$\chi_{A_{1}A_{2}}(\mathbf{r})\otimes\chi_{B_{1}B_{2}}(\mathbf{r})$,
where $\mathbf{r}$ characterizes the displacement, which represent
products of local Gaussian trace-preserving operations
\cite{Lindblad_00,Caruso_08},
$\mathcal{E}_{A}(\mathbf{r})\otimes\mathcal{E}_{B}(\mathbf{r})$.
Consequently, we can take the purification
$|\chi\rangle_{A_{1}B_{1}A_{2}B_{2}E_{\chi}}$ in the form for
which there is a measurement with CM $\tilde{\Gamma}_{E_{\chi}}$
on subsystem $E_{\chi}$, which projects it onto the product
states.

Let $\gamma_{\pi}^{k}$ and $\Gamma_{j_{k}}$, $j=A,B,E$, now denote
CMs of the discussed purification and optimal measurements for the
state $\rho_{A_{k}B_{k}}$, $k=\mathrm{in},\mathrm{out}$, i.e.,
$E_{\downarrow}^{G}(\rho_{{A}_{k}{B}_{k}})=f(\gamma_{\pi}^{k},\Gamma_{A_k},\Gamma_{B_k},\Gamma_{E_k})$.
The inequality (\ref{xmonotonicity}) is then a consequence of the
following chain of inequalities:
\begin{eqnarray*}\label{xIneqs}
E_{\downarrow}^{G}(\rho_{{A}_{\rm out}{B}_{\rm out}})&\leq&
f(\gamma_{\pi}^{{\rm out}},\Gamma_{A_{\rm out}},\Gamma_{B_{\rm
out}},\tilde{\Gamma}_{E_{\chi}}\oplus\Gamma_{E_{\rm
in}})\nonumber\\
&\leq&f(\gamma_{\pi}^{{\rm in}},\tilde{\Gamma}_{A_{\rm
in}},\tilde{\Gamma}_{B_{\rm in}},\Gamma_{E_{\rm in}})\leq
E_{\downarrow}^{G}(\rho_{A_{\rm in}B_{\rm in}}).
\end{eqnarray*}
The first inequality is satisfied because $f$ cannot decrease by
replacing the optimal measurement having CM $\Gamma_{E_{\rm out}}$
by a (generally suboptimal) product measurement with CM
$\tilde{\Gamma}_{E_{\chi}}\oplus\Gamma_{E_{\rm in}}$. Next, the
latter measurement projects $|\chi\rangle$ onto a product state
and $|\Psi\rangle_{A_{\rm in}B_{\rm in}E_{\rm in}}$ onto the state
$\rho_{A_{\rm in}B_{\rm in}|E_{\rm in}}$, and hence the right-hand
side of the first inequality is equal to the mutual information of
outcomes of measurements with CMs $\Gamma_{A_{\rm out}}$ and
$\Gamma_{B_{\rm out}}$ on the state
$(\mathcal{E}_{A}\otimes\mathcal{E}_{B})(\rho_{A_{\rm in}B_{\rm
in}|E_{\rm in}})$, where $\mathcal{E}_{A,B}$ are operations with
zero displacements due to the independence of mutual information
from displacements. Since operation $\mathcal{E}_{j}$, $j=A,B$, is
trace-preserving, it can be realized by a Gaussian unitary
operation $U_{j}$ on input system $j_{\rm in}$ and vacuum ancilla
$j_{\rm anc}$, followed by discarding of a part of output system,
$j_{\rm disc}$, and adding classical Gaussian noise
\cite{Eisert_03} (see Fig.~\ref{fig_1}(b)). The noise can be
integrated into a new measurement with CM $\bar{\Gamma}_{j_{\rm
out}}$ and we can also work with a measurement on a larger system
$(j_{\rm out}j_{\rm disc})$ with CM $\bar{\Gamma}_{j_{\rm
out}}\oplus\Gamma_{j_{\rm disc}}$, because it never gives a
smaller mutual information than the original measurement
\cite{Cover_06}. Moreover, unitary $U_{j}$ can be integrated into
a new measurement with CM $\Gamma_{j_{\rm in}j_{\rm anc}}$ on
system $(j_{\rm in}j_{\rm anc})$, which yields the same mutual
information which can be further rewritten in terms of some
measurements with CMs $\tilde{\Gamma}_{A_{\rm in}}$ and
$\tilde{\Gamma}_{B_{\rm in}}$ on state $\rho_{A_{\rm in}B_{\rm
in}|E_{\rm in}}$ \cite{Mista_15}. Therefore, the second inequality
holds. Finally, CMs $\tilde{\Gamma}_{A_{\rm in}}$ and
$\tilde{\Gamma}_{B_{\rm in}}$ cannot give a larger mutual
information than optimal CMs $\Gamma_{A_{\rm in}}$ and
$\Gamma_{B_{\rm in}}$, and thus the last inequality is fulfilled,
which completes the monotonicity proof.

{\it Computability}.---GIE can be calculated analytically for
several classes of two-mode Gaussian states. Without loss of
generality \cite{Mista_15} we can take CMs of the states in the
standard form \cite{Simon_00,Giedke_03a}
\begin{eqnarray}\label{xgammasymst}
\gamma_{AB}=\left(\begin{array}{cc}
a\openone & \kappa \\
\kappa & b\openone \\
\end{array}\right)
\end{eqnarray}
with $\kappa=\mbox{diag}(k_{x},-k_{p})$, where $k_x\geq k_p\geq0$.
We evaluate GIE both for symmetric states with $a=b$ as well as
for some asymmetric states. First, we calculate an easier
computable upper bound on GIE obtained by reversing the order of
optimization in its definition (\ref{xEGdownarrow}). Next, we find
for some fixed measurements on modes $A$ and $B$ a measurement on
subsystem $E$ giving minimal $f$ which at the same time saturates
the bound (see Appendix~\cite{SI} for details). It turns out, that for
all symmetric (asymmetric) states considered here, GIE is achieved
by double homodyne detection on modes $A$ and $B$, and homodyne
detection (heterodyne detection, i.e., projection onto coherent
states) on subsystem $E$. We have found GIE for the following
three sets of states:

(i) {\it Symmetric GLEMS} \cite{Adesso_04}.--- The states
$(\equiv\rho_{AB}^{(1)})$ have one unit symplectic eigenvalue
\cite{Williamson_36} whence $k_x=a-1/(a+k_p)$ and a subsystem $E$
is single-mode. For all the states GIE reads as \cite{SI}
\begin{equation}\label{xGIEsymGLEMS}
E_{\downarrow}^{G}\left(\rho_{AB}^{(1)}\right)=\ln\left(\frac{a}{\sqrt{a^2-k_p^2}}\right).
\end{equation}
If $k_{x}=k_{p}$ symmetric GLEMS satisfy $a^2-k_{p}^2=1$ and
therefore they reduce to pure states $(\equiv\rho_{AB}^{p})$.
Equation (\ref{xGIEsymGLEMS}) then gives
$E_{\downarrow}^{G}(\rho_{AB}^{p})=\ln(a)$ \cite{Mista_15}.

(ii) {\it Symmetric squeezed thermal states}
\cite{Botero_03}.---The states $(\equiv\rho_{AB}^{(2)})$ fulfil the
condition $k_{x}=k_{p}\equiv k$ and they are entangled iff $a-k<1$ \cite{Simon_00,Giedke_03a}.
For all the entangled states which satisfy $a\leq2.41$ GIE is equal to \cite{SI}
\begin{equation}\label{xGIEisfinal}
E_{\downarrow}^{G}\left(\rho_{AB}^{(2)}\right)=\ln\left[\frac{(a-k)^2+1}{2(a-k)}\right],
\end{equation}
whereas for separable states
$E_{\downarrow}^{G}(\rho_{AB}^{(2)})=0$ by faithfulness.

(iii) {\it Asymmetric squeezed thermal GLEMS}.---The states
$(\equiv\rho_{AB}^{(3)})$ fulfill the condition $k_{x}=k_{p}\equiv
k$ and possess one unit symplectic eigenvalue. For all the states
for which $\sqrt{ab}\leq2.41$ GIE is given by \cite{SI}
\begin{equation}\label{xGIEGLEMSfinal}
E_{\downarrow}^{G}\left(\rho_{AB}^{(3)}\right)=\ln\left(\frac{a+b}{|a-b|+2}\right).
\end{equation}

{\it Discussion and conclusions}.--- We have proposed a quantifier
of Gaussian entanglement GIE which compromises between
computability and operational significance. Closed formulae for
GIE for two classes of symmetric states have been obtained,
Eqs.~(\ref{xGIEsymGLEMS}) and (\ref{xGIEisfinal}), which can be
compactly written as
$E_{\downarrow}^{G}(\rho_{AB})=\ln\{[\tilde{\nu}_{-}+(\tilde{\nu}_{-})^{-1}]/2\}$
if $\tilde{\nu}_{-}<1$ and $E_{\downarrow}^{G}(\rho_{AB})=0$ if
$\tilde{\nu}_{-}\geq1$, where
$\tilde{\nu}_{-}=\sqrt{(a-k_{x})(a-k_{p})}$. Interestingly, this
is nothing but GR2 entanglement for symmetric states
\cite{Adesso_12,Giedke_03a}. As the GIE for some asymmetric
states, Eq.~(\ref{xGIEGLEMSfinal}), also coincides with the GR2
entanglement \cite{SI} we conjecture, that the two quantities are
equivalent. The confirmation or refutation of the conjecture as
well as analysis of the other properties of GIE is left for future
research. We hope that the present results will stimulate further
studies of physically meaningful computable entanglement measures.

We would like to thank J. Fiur\'a\v{s}ek and G. Adesso for
fruitful discussions. L. M. acknowledges the Project No.
P205/12/0694 of GA\v{C}R.

\clearpage
\newpage
\pagenumbering{arabic}
\onecolumngrid
\begin{center}
Supplementary Information

\vspace{0.2cm}

\textbf{Gaussian intrinsic entanglement}

\vspace{0.2cm}

Ladislav Mi\v{s}ta, Jr. and Richard Tatham

\vspace{0.4cm}

\end{center}
\twocolumngrid
\appendix
\section{GIE for two-mode Gaussian states}\label{Sec_0}

For a Gaussian state $\rho_{AB}$ of two modes $A$ and $B$ GIE is
defined explicitly as
\begin{equation}\label{GIE}
E_{\downarrow}^{G}\left(\rho_{AB}\right)=\mathop{\mbox{sup}}_{\Gamma_{A},\Gamma_{B}}
\mathop{\mbox{inf}}_{\Gamma_{E}}f\left(\gamma_{\pi},\Gamma_{A},\Gamma_{B},\Gamma_{E}\right),
\end{equation}
where
\begin{equation}\label{f}
f\left(\gamma_{\pi},\Gamma_{A},\Gamma_{B},\Gamma_{E}\right)=\frac{1}{2}\ln\left(\frac{\mbox{det}\sigma_{A}\mbox{det}\sigma_{B}}{\mbox{det}\sigma_{AB}}\right)
\end{equation}
with
\begin{eqnarray}\label{sigma}
\sigma_{AB}&=&\gamma_{AB|E}+\Gamma_{A}\oplus\Gamma_{B},
\end{eqnarray}
where $\sigma_{A,B}$ are local submatrices of $\sigma_{AB}$ and
$\Gamma_{A}$ and $\Gamma_{B}$ are single-mode CMs of Gaussian
measurements on modes $A$ and $B$, respectively. Here,
\begin{eqnarray}\label{gammacond}
\gamma_{AB|E}&=&\gamma_{AB}-\gamma_{ABE}\frac{1}{\gamma_{E}+\Gamma_{E}}\gamma_{ABE}^{T}
\end{eqnarray}
is a CM of a conditional state $\rho_{AB|E}$ \cite{Giedke_02} of
modes $A$ and $B$ obtained by a Gaussian measurement with CM
$\Gamma_E$ on purifying subsystem $E$ of the minimal purification
\cite{Caruso_11} of the state $\rho_{AB}$, where $\gamma_{AB}$ is
a CM of the state $\rho_{AB}$ and $\gamma_{ABE}$ and $\gamma_{E}$
denote the blocks of the CM of the purification
$(\equiv\gamma_{\pi})$ expressed with respect to the $AB|E$
splitting, i.e.,
\begin{eqnarray}\label{gammapi}
\gamma_{\pi}=\left(\begin{array}{cc}
\gamma_{AB} & \gamma_{ABE} \\
\gamma_{ABE}^{T} & \gamma_{E} \\
\end{array}\right).
\end{eqnarray}

If the Gaussian state $\rho_{AB}$ is a pure state
($\equiv\rho^{p}$), the off-diagonal block $\gamma_{ABE}$ is a
zero matrix and the GIE then can be calculated easily. In the main
text as well as in Ref.~\cite{Mista_15} it was shown, that in this
case the GIE coincides with the Gaussian R\'enyi-2 (GR2)
entanglement ($\equiv E_{2}^{G}(\rho^{p})$) \cite{Adesso_12},
\begin{equation}\label{GIEpure}
E_{\downarrow}^{G}\left(\rho^{p}\right)=E_{2}^{G}\left(\rho^{p}\right)=\frac{1}{2}\ln\left(\mbox{det}\gamma_{A}\right),
\end{equation}
where $\gamma_{A}$ is a CM of the reduced state of mode $A$ of the
state $\rho^{p}$.

In what follows, we focus on calculation of GIE for mixed two-mode Gaussian states. The states possess the blocks $\gamma_{ABE}$ and
$\gamma_{E}$ of the form \cite{Mista_15}
\begin{eqnarray}\label{gammaABEgammaE}
\gamma_{ABE}=S^{-1}\gamma_{ABE}^{(0)},\quad\gamma_{E}=\gamma_{E}^{(0)},
\end{eqnarray}
where
\begin{eqnarray}\label{gammaABEgammaE0}
\gamma_{ABE}^{(0)}=\left(\begin{array}{c}
\bigoplus_{i=1}^{R}\sqrt{\nu_{i}^{2}-1}\sigma_{z} \\
\mathbb{O}_{2(2-R)\times2R} \\
\end{array}\right),\quad\gamma_{E}^{(0)}=\bigoplus_{i=1}^{R}\nu_{i}\openone.
\end{eqnarray}
Here $\sigma_{z}=\mbox{diag}(1,-1)$ is the diagonal Pauli-$z$
matrix, $\mathbb{O}_{I\times J}$ is the $I\times J$ zero matrix,
$\openone$ is the $2\times 2$ identity matrix and $S$ is a
symplectic matrix, i.e., a $4\times4$ real matrix satisfying the
symplectic condition
\begin{equation}\label{Scondition}
S\Omega S^{T}=\Omega,
\end{equation}
where
\begin{equation}\label{Omega}
\Omega=\bigoplus_{i=1}^{2}J,\quad J=\left(\begin{array}{cc}
0 & 1 \\
-1 & 0\\
\end{array}\right),
\end{equation}
that brings the CM $\gamma_{AB}$ to the Williamson normal form
\cite{Williamson_36}
\begin{equation}\label{Williamson}
S\gamma_{AB}S^{T}=\mbox{diag}(\nu_{1},\nu_{1},\nu_{2},\nu_{2}).
\end{equation}
Here $\nu_{1}\geq\nu_{2}\geq1$ are the so called symplectic
eigenvalues of CM $\gamma_{AB}$ and $R=1,2$ is the number of the
symplectic eigenvalues strictly greater than one. The use of
Eq.~(\ref{gammaABEgammaE}) on the right-hand side (RHS) of
Eq.~(\ref{gammacond}) further yields for the CM $\gamma_{AB|E}$
the expression
\begin{equation}\label{gammacondS}
\gamma_{AB|E}=S^{-1}\gamma_{AB|E}^{(0)}(S^{-1})^{T}
\end{equation}
with
\begin{eqnarray}\label{gammacond0}
\gamma_{AB|E}^{(0)}&=&\gamma_{AB}^{(0)}-\gamma_{ABE}^{(0)}\frac{1}{\gamma_{E}^{(0)}+\Gamma_{E}}(\gamma_{ABE}^{(0)})^{T},
\end{eqnarray}
where $\gamma_{AB}^{(0)}$ denotes the Williamson normal form
(\ref{Williamson}) of CM $\gamma_{AB}$, i.e.,
$\gamma_{AB}^{(0)}=\mbox{diag}(\nu_{1},\nu_{1},\nu_{2},\nu_{2})$.

In order to calculate the GIE we now need to express CM
(\ref{gammacond0}) as well as the symplectic matrix $S$ appearing
in Eq.~(\ref{gammacondS}) in terms of the elements of the CM
$\gamma_{AB}$. For this purpose it is convenient to express the CM
in a block form with respect to the $A|B$ splitting,
\begin{eqnarray}\label{gammasblock}
\gamma_{AB}=\left(\begin{array}{cc}
A & C \\
C^{T} & B \\
\end{array}\right).
\end{eqnarray}
Owing to the invariance of GIE (\ref{GIE}) under the Gaussian
local unitary operations \cite{Mista_15} we can without loss of
generality assume CM (\ref{gammasblock}) to be in the standard
form \cite{Simon_00},
\begin{eqnarray}\label{gammast1}
\gamma_{AB}=\left(\begin{array}{cccc}
a & 0 & c_x & 0\\
0 & a & 0 & c_p \\
c_x & 0 & b & 0 \\
0 & c_p & 0 & b \\
\end{array}\right)
\end{eqnarray}
with $c_{x}\geq|c_{p}|\geq0$. Since states with $c_{x}c_{p}\geq0$
are separable \cite{Simon_00} and thus possess zero GIE
\cite{Mista_15}, in calculations we can restrict ourself only to
CMs satisfying $c_{x}c_{p}<0$. Introducing new more convenient
parameters $k_{x}\equiv c_{x}$ and $k_{p}\equiv|c_{p}|=-c_{p}$, we
arrive at the following standard-form CM which we shall consider
in what follows \cite{Giedke_03a}:
\begin{eqnarray}\label{gammast2}
\gamma_{AB}=\left(\begin{array}{cccc}
a & 0 & k_x & 0\\
0 & a & 0 & -k_p \\
k_x & 0 & b & 0 \\
0 & -k_p & 0 & b \\
\end{array}\right),
\end{eqnarray}
where $k_x\geq k_p>0$.

The symplectic eigenvalues of CM (\ref{gammast2}) can be
calculated conveniently from the eigenvalues of the matrix
$i\Omega\gamma_{AB}$ which are of the form $\{\pm \nu_{1},\pm
\nu_2\}$ \cite{Vidal_02}. In terms of parameters $a,b,k_{x}$ and
$k_{p}$ they read explicitly as
\begin{equation}\label{nu12}
\nu_{1,2}=\sqrt{\frac{\Delta\pm\sqrt{D}}{2}},
\end{equation}
where
\begin{eqnarray}\label{DeltaD}
\Delta&=&a^2+b^2-2k_xk_p,\nonumber\\
D&=&\Delta^2-4\mbox{det}\gamma_{AB}\nonumber\\
&=&\left(a^2-b^2\right)^2+4\left(ak_x-bk_p\right)\left(bk_x-ak_p\right).\nonumber\\
\end{eqnarray}

Similarly, we can express the symplectic matrix $S$ which brings
CM (\ref{gammast2}) to Williamson normal form (\ref{Williamson})
in terms of parameters $a,b,k_{x}$ and $k_{p}$. This can be done
using either a method of Ref.~\cite{Serafini_05} or a method of
Ref.~\cite{Pirandola_09}. For a generic two-mode CM
(\ref{gammast2}) the form of the matrix $S$ is complex and
therefore we do not write it here explicitly. In what follows, we
work with particular subclasses of the class of generic two-mode
Gaussian states for which $S$ attains a simple form which is
presented explicitly in the respective subsection.

\section{GIE for symmetric states}\label{Sec_I}

In this section we calculate GIE defined in Eq.~(8) of the main
text for some subclasses of the class of two-mode symmetric
Gaussian states. The states are characterized by the condition
$a=b$ whence their standard-form CMs (\ref{gammast1}) and
(\ref{gammast2}) reduce to
\begin{eqnarray}\label{gammastsym}
\gamma_{AB}=\left(\begin{array}{cccc}
a & 0 & c_x & 0\\
0 & a & 0 & c_p \\
c_x & 0 & a & 0 \\
0 & c_p & 0 & a \\
\end{array}\right)
\end{eqnarray}
and
\begin{eqnarray}\label{gammasymst}
\gamma_{AB}=\left(\begin{array}{cccc}
a & 0 & k_x & 0\\
0 & a & 0 & -k_p \\
k_x & 0 & a & 0 \\
0 & -k_p & 0 & a \\
\end{array}\right),
\end{eqnarray}
respectively. For the sake of further use let us also recap here
that the matrix (\ref{gammasymst}) describes a CM of a physical
quantum state if and only if $a^2-k_{x}^{2}\geq 1$ and that the CM
corresponds to an entangled state if and only if
$1>(a-k_{x})(a-k_{p})$ \cite{Giedke_03a}.

From Eq.~(\ref{nu12}) it further follows that the symplectic
eigenvalues of CM (\ref{gammasymst}) read explicitly as
\begin{eqnarray}\label{nu12sym}
\nu_{1}&=&\sqrt{\left(a+k_x\right)\left(a-k_p\right)},\nonumber\\
\nu_{2}&=&\sqrt{\left(a-k_x\right)\left(a+k_p\right)}.
\end{eqnarray}

As for the symplectic matrix $S$, we calculate it using the method
of Ref.~\cite{Serafini_05}. Here, we seek the matrix in the form
of a product $S=\left(\oplus_{i=1}^{2}V^{\ast}\right)W^{T}$, where
\begin{equation}\label{U}
V=\frac{1}{\sqrt{2}}\left(\begin{array}{cc}
i & -i \\
1 & 1\\
\end{array}\right)
\end{equation}
and $W$ contains in its columns the eigenvectors of the matrix
$i\Omega\gamma_{AB}$ which are chosen such that $S$ is real, it
satisfies the symplectic condition (\ref{Scondition}) and it does
not mix position and momentum quadratures. Thus we find the
symplectic matrix $S$ that brings the CM (\ref{gammasymst}) to the
Williamson normal form (\ref{Williamson}) to be the following
product
\begin{equation}\label{Ssym}
S=(S_{A}\oplus S_{B})U_{BS}.
\end{equation}
Here,
\begin{equation}\label{Ubeamsplitter}
U_{BS}=\frac{1}{\sqrt{2}}\left(\begin{array}{cc}
\openone & \openone \\
-\openone & \openone\\
\end{array}\right)
\end{equation}
is a matrix describing a balanced beam splitter and
\begin{equation}\label{SAB}
S_{A}=\left(\begin{array}{cc}
z_{A}^{-1} & 0 \\
0 & z_{A}\\
\end{array}\right),\quad S_{B}=\left(\begin{array}{cc}
z_{B} & 0 \\
0 & z_{B}^{-1}\\
\end{array}\right)
\end{equation}
with $z_{A}=\sqrt[4]{\frac{a+k_x}{a-k_p}}>1$ and
$z_{B}=\sqrt[4]{\frac{a+k_p}{a-k_x}}>1$ are matrices corresponding
to local squeezing transformations in quadratures $x_{A}$ and
$p_{B}$, respectively.

Making use of the symplectic eigenvalues (\ref{nu12sym}) and the
symplectic matrix (\ref{Ssym}) we can now express CM
(\ref{gammacondS}) solely in terms of the parameters $a,k_x$ and
$k_p$ and the CM $\Gamma_{E}$. However, this form of CM
(\ref{gammacondS}) is not suitable for analytical calculation of
the GIE, because the CM depends on CM $\Gamma_{E}$ that we
minimize over in the definition of GIE, Eq.~(\ref{GIE}), in a
complicated way via the term
$\frac{1}{\gamma_{E}^{(0)}+\Gamma_{E}}$. Although the term can be
calculated explicitly for an arbitrary two-mode Gaussian state,
the obtained form of CM (\ref{gammacond0}) still depends on
elements of CM $\Gamma_{E}$ in a way which is too complicated for
analytical calculations. Nevertheless, there are some subclasses
of the set of symmetric states for which the CM simplifies such
that the GIE can be calculated analytically. These states can be
identified if one realizes that the dimension of CM
$\gamma_{E}^{(0)}$ and therefore also CM $\Gamma_{E}$ is
determined by the number $R$, which appears in
Eq.~(\ref{gammaABEgammaE0}) and which denotes the number of
symplectic eigenvalues of a respective two-mode Gaussian state
which are strictly greater than one. Therefore, the set of mixed
two-mode Gaussian states splits into two subsets containing states
with $R=1$ and $R=2$, respectively. States with $R=1$ are more
simple because for them the matrix $\gamma_{E}^{(0)}+\Gamma_{E}$
is just single-mode and therefore the inverse
$\frac{1}{\gamma_{E}^{(0)}+\Gamma_{E}}$ is simple. In the
following subsection we show, that for symmetric states with $R=1$
the GIE can be calculated analytically. The next subsection is
then dedicated to analytical evaluation of GIE for a subclass of
more complicated symmetric states with $R=2$. Specifically, in
this subsection we calculate GIE for a subclass of the symmetric
states satisfying $\nu_1=\nu_{2}$ \cite{Botero_03}, which are the
so called symmetric squeezed thermal states.

\subsection{GIE for symmetric GLEMS}\label{subsec_symGLEMS}

Let us consider a two-mode Gaussian state $\rho_{AB}^{(1)}$
with a CM $\gamma_{AB}^{(1)}$ and $R=1$. Since $\nu_{2}=1$ holds for
this state it is a Gaussian state with
a partial minimal uncertainty. This state is also known to be the
Gaussian least entangled state for given global and local purities
(GLEMS) \cite{Adesso_04} and it possesses the other symplectic
eigenvalue equal to
\begin{equation}\label{GLEMSnu}
\nu\equiv\nu_{1}=\sqrt{\mbox{det}\gamma_{AB}^{(1)}}
\end{equation}
as can be seen from Eq.~(\ref{Williamson}). Because $R=1$, the
minimal purification of GLEMS contains only a single purifying
mode $E$ and therefore also the CM $\Gamma_{E}$ appearing in CM
(\ref{gammacond0}) is single-mode.

Let us assume now the CM in the form
$\Gamma_E=P(\varphi)\mbox{diag}(V_{x},V_{p})P^{T}(\varphi)$, where
\begin{equation}\label{UV}
P(\varphi)= \left(
\begin{array}{cc}
\cos\varphi  & -\sin \varphi \\
\sin\varphi & \cos\varphi \\
\end{array}
\right)
\end{equation}
with $\varphi\in[0,\pi)$, $V_{x}=\tau e^{2t}$ and $V_{p}=\tau
e^{-2t}$, where $\tau\geq1$ and $t\geq0$. Using relation
$P^{T}(\varphi)=\sigma_{z}P(\varphi)\sigma_{z}$ one gets after
some algebra that the CM (\ref{gammacondS}) can be expressed as
\begin{eqnarray}\label{gammacondS2}
\gamma_{AB|E}^{(1)}=S^{-1}\left[\gamma_{A|E}(\nu)\oplus\openone_{B}\right](S^{-1})^{T}.
\end{eqnarray}
\begin{figure}[t]
\includegraphics[width=0.8\columnwidth]{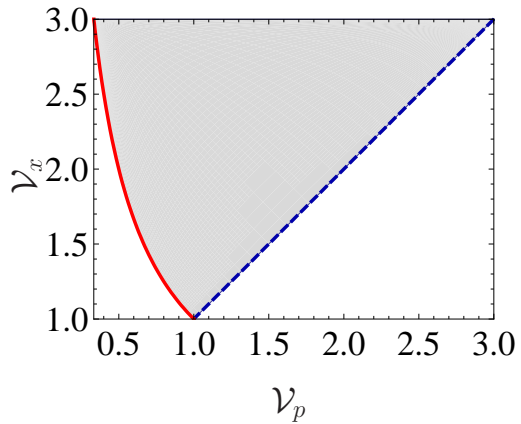}
\caption{(Color online) An example of the set $\mathscr{M}$ (gray
area) for $\nu=3$. Solid red curve and dashed blue curve depict
the boundary curves $\mathcal{V}_{x}=1/\mathcal{V}_{p}$ and
$\mathcal{V}_{x}=\mathcal{V}_{p}$, respectively.}
\label{fig_region}
\end{figure}
Here,
\begin{equation}\label{gammaAE}
\gamma_{A|E}(\nu)=P^{T}(\varphi)\mbox{diag}(\mathcal{V}_{x},\mathcal{V}_{p})P(\varphi),
\end{equation}
where
\begin{equation}\label{calVxVp}
\mathcal{V}_{x}=\frac{\nu V_{x}+1}{\nu+V_{x}},\quad
\mathcal{V}_{p}=\frac{\nu V_{p}+1}{\nu+V_{p}}
\end{equation}
are the eigenvalues of the CM. From the definition of parameters
$V_{x}$ and $V_{p}$ given below Eq.~(\ref{UV}) it then follows
that $V_{x}\geq1$, $V_{x}\geq V_{p}\geq0$ and $V_{x}V_{p}\geq 1$.
Consequently, at given $\nu$ the eigenvalues (\ref{calVxVp}) lie
in the subset $\mathscr{M}$ of the
$(\mathcal{V}_{p},\mathcal{V}_{x})$-plane such that if
$\mathcal{V}_{p}\in[1/\nu,1]$ then
$\mathcal{V}_{x}\in[1/\mathcal{V}_{p},\nu]$, whereas if
$\mathcal{V}_{p}\in(1,\nu]$ then
$\mathcal{V}_{x}\in[\mathcal{V}_{p},\nu]$ (see
Fig.~\ref{fig_region}).

For symmetric GLEMS we can further insert for the symplectic
matrix $S$ on the RHS of Eq.~(\ref{gammacondS2}) the decomposition (\ref{Ssym})
which yields
\begin{eqnarray}\label{gammacond1}
\gamma_{AB|E}^{(1)}=U_{BS}^{T}\left[S_{A}^{-1}\gamma_{A|E}(\nu)(S_{A}^{-1})^{T}\oplus\gamma_{B}^{\rm
sq}\right]U_{BS},
\end{eqnarray}
where $\gamma_{B}^{\rm
sq}=S_{B}^{-1}(S_{B}^{-1})^{T}=\mbox{diag}(z_{B}^{-2},z_{B}^{2})$,
and where we have used the orthogonality of the matrix
(\ref{Ubeamsplitter}).

The form of CM (\ref{gammacond1}) is now suitable for carrying out
of the last step of calculation of GIE, Eq.~(\ref{GIE}), which is
the optimization of the function (\ref{f}) with respect to CMs
$\Gamma_{A}, \Gamma_{B}$ and $\Gamma_{E}$. Here, we perform the
optimization by means of the method, which was used in
Ref.~\cite{Mista_15} for calculation of the GIE for a particular
instance of symmetric GLEMS given by the two-mode reduction of the
three-mode continuous-variable (CV) Greenberger-Horne-Zeilinger
(GHZ) state \cite{Loock_00}. In this method, we calculate the GIE
for a Gaussian state $\rho_{AB}^{(1)}$ in two steps. First, we
calculate an easier computable upper bound
\begin{equation}\label{Ubound}
U\left(\rho_{AB}^{(1)}\right)\equiv\mathop{\mbox{inf}}_{\Gamma_{E}}\mathop{\mbox{sup}}_{\Gamma_{A},\Gamma_{B}}
f(\gamma_{\pi},\Gamma_{A},\Gamma_{B},\Gamma_{E})
\end{equation}
on GIE, $E_{\downarrow}^{G}(\rho_{AB}^{(1)})\leq
U(\rho_{AB}^{(1)})$. In the second step, we find for particular
fixed CMs $\Gamma_{A}$ and $\Gamma_{B}$ an optimal CM
$\tilde{\Gamma}_{E}$ which minimizes
$f(\gamma_{\pi},\Gamma_{A},\Gamma_{B},\Gamma_{E})$, i.e.,
$f(\gamma_{\pi},\Gamma_{A},\Gamma_{B},\tilde{\Gamma}_{E})=\mathop{\mbox{inf}}_{\Gamma_{E}}f(\gamma_{\pi},\Gamma_{A},\Gamma_{B},\Gamma_{E})$,
and which at the same time saturates the upper bound, i.e.,
$U(\rho_{AB}^{(1)})=f(\gamma_{\pi},\Gamma_{A},\Gamma_{B},\tilde{\Gamma}_{E})$.
As a consequence, the GIE for state $\rho_{AB}^{(1)}$ then reads
as
\begin{equation}\label{GIE2}
E_{\downarrow}^{G}\left(\rho_{AB}^{(1)}\right)=f(\gamma_{\pi},\Gamma_{A},\Gamma_{B},\tilde{\Gamma}_{E}).
\end{equation}

In order to calculate the upper bound (\ref{Ubound}), we first
calculate the quantity
\begin{equation}\label{IcG}
\mathcal{I}_{c}^{G}\left(\rho_{AB|E}^{(1)}\right)\equiv\mathop{\mbox{sup}}_{\Gamma_{A},\Gamma_{B}}
f(\gamma_{\pi},\Gamma_{A},\Gamma_{B},\Gamma_{E}),
\end{equation}
which is the Gaussian classical mutual information (GCMI) of the
conditional quantum state $\rho_{AB|E}^{(1)}$ with CM
(\ref{gammacond1}) \cite{Mista_11}. Substituting on the RHS of
Eq.~(\ref{gammacond1}) for the matrix $U_{BS}$ from
Eq.~(\ref{Ubeamsplitter}) one finds, that the block form with
respect to the $A|B$ splitting of CM (\ref{gammacond1}) reads as
\begin{eqnarray}\label{gammasblock1}
\gamma_{AB|E}^{(1)}=\left(\begin{array}{cc}
A & C \\
C & A \\
\end{array}\right)
\end{eqnarray}
and therefore the CM is symmetric under the exchange of modes $A$
and $B$. This implies, that the standard form of CM
(\ref{gammacond1}) attains the symmetric form (\ref{gammastsym}).
Denoting now the parameters of the standard form as
$\tilde{a},\tilde{c}_{x}$ and $\tilde{c}_{p}$ we can use the
result of Ref.~\cite{Mista_15} to calculate the quantity
(\ref{IcG}). Specifically, in Ref.~\cite{Mista_15} it was shown,
that for states with the symmetric standard-form CM where the
parameters $\tilde{a}$ and $\tilde{c}_{x}$ satisfy inequality
\begin{equation}\label{IcGcondition}
2+\frac{1}{\tilde{a}}-\tilde{s}\geq0,
\end{equation}
where $\tilde{s}\equiv\sqrt{\tilde{a}^{2}-\tilde{c}_{x}^2}$, the
GCMI (\ref{IcG}) is attained by homodyne detections of quadratures
$x_{A}$ and $x_{B}$ on modes $A$ and $B$, for which it is of the
form:
\begin{equation}\label{IcGsym}
\mathcal{I}_{c}^{G}\left(\rho_{AB|E}^{(1)}\right)=\frac{1}{2}\ln\frac{\tilde{a}^2}{\tilde{a}^2-\tilde{c}_{x}^2}=-\ln\sqrt{1-g^2},
\end{equation}
where $g\equiv\tilde{c}_{x}/\tilde{a}$. By calculating explicitly
the blocks $A$ and $C$ of CM (\ref{gammasblock1}) we can express
the quantities $\tilde{a}$ and $g$ in terms of the parameters
$a,k_x$ and $k_p$ of the original state and the variables
$\mathcal{V}_{p},\mathcal{V}_{x}$ and $\varphi$ over which we
carry out minimization. Making use of the formula
$\tilde{a}=\sqrt{\mbox{det}A}$ one finds that
\begin{eqnarray}\label{tildea}
\tilde{a}=\frac{\sqrt{1+\mathcal{V}_{x}\mathcal{V}_{p}+2[\mathcal{V}_{+}\cosh(2q)+\mathcal{V}_{-}\sinh(2q)\cos(2\varphi)]}}{2},\nonumber\\
\end{eqnarray}
where $\mathcal{V}_{\pm}=(\mathcal{V}_{x}\pm\mathcal{V}_{p})/2$
and $q=\ln(z_{A}z_{B})$, whereas for the quantity $g$ one gets
\cite{Mista_15}
\begin{eqnarray}\label{g}
g=\frac{\mathcal{K}}{\tilde{a}^2}+\sqrt{\left(\frac{\mathcal{K}}{\tilde{a}^2}-1\right)^2-\frac{1}{\tilde{a}^2}}
\end{eqnarray}
with $\mathcal{K}=(\mathcal{V}_{x}\mathcal{V}_{p}-1)/4$.

For calculation of the upper bound (\ref{Ubound}) it remains to
minimize the RHS of Eq.~(\ref{IcGsym}) over all 3-tuples
$(\mathcal{V}_{p},\mathcal{V}_{x},\varphi)$ belonging to the
Cartesian product $\mathscr{O}=\mathscr{M}\times[0,\pi)$, where
the set $\mathscr{M}$ is defined below equation (\ref{calVxVp}).
This amounts to the minimization of the RHS of Eq.~(\ref{g}) on
the set $\mathscr{O}$, which can be performed exactly as in
Ref.~\cite{Mista_15}. Namely, comparison of the quantities
(\ref{tildea}) and (\ref{g}) with the same quantities for the CV
GHZ state analyzed in Ref.~\cite{Mista_15} reveals, that they are
exactly the same. The only difference is in the parameter $q$
which is for the present case defined below Eq.~(\ref{tildea}),
whereas for the CV GHZ state it is equal to
$q=r+\ln(\sqrt{x_{-}/x_{+}})/2$ with $x_{\pm}=(e^{\pm 2r}+2e^{\mp
2r})/3$, where $r\geq0$ is a squeezing parameter. Minimizing
therefore the quantity $g$, Eq.~(\ref{g}), exactly as in
Ref.~\cite{Mista_15} and restricting ourselves only to entangled
states satisfying the necessary and sufficient condition for
entanglement, $1>(a-k_{x})(a-k_{p})$, one finds, that the quantity
possesses three candidates for the minimum which lie on the edges
of the volume $\mathscr{O}$ and correspond to:

1. Homodyne detection of quadrature $p_{E}$ on mode $E$, i.e.,
$\Gamma_{E}=\Gamma_{p}^{t\rightarrow+\infty}$, where
$\Gamma_{p}^{t}\equiv\mbox{diag}(e^{2t},e^{-2t})$, or discarding
of mode $E$, which yields
\begin{eqnarray}\label{U1}
U_{1}\left(\rho_{AB}^{(1)}\right)=\ln\left(a\sqrt{\frac{a+k_p}{a+k_x}}\right)=\ln\left(\frac{a}{\sqrt{a^2-k_x^2}}\right).\nonumber\\
\end{eqnarray}
Here, the second equality is a consequence of the equality
\begin{equation}\label{kxkp}
a+k_p=\frac{1}{a-k_x},
\end{equation}
which follows from the defining equality for symmetric GLEMS,
$\nu_{2}=\sqrt{(a-k_x)(a+k_p)}=1$.

2. Heterodyne detection on mode $E$, i.e., $\Gamma_{E}=\openone$,
which gives
\begin{eqnarray}\label{U2}
U_{2}\left(\rho_{AB}^{(1)}\right)=\ln\left[\frac{1}{2}\left(z_{A}z_{B}+\frac{1}{z_{A}z_{B}}\right)\right],
\end{eqnarray}
where the parameters $z_{A}$ and $z_{B}$ are defined below Eq.~(\ref{SAB}).

3. Homodyne detection of quadrature $x_{E}$ on mode $E$, i.e.,
$\Gamma_{E}=\Gamma_{x}^{t\rightarrow+\infty}$, where
$\Gamma_{x}^{t}\equiv\mbox{diag}(e^{-2t},e^{2t})$, for which the
upper bound (\ref{Ubound}) is equal to
\begin{eqnarray}\label{U3}
U_{3}\left(\rho_{AB}^{(1)}\right)=\ln\left(a\sqrt{\frac{a-k_x}{a-k_p}}\right)=\ln\left(\frac{a}{\sqrt{a^2-k_p^2}}\right).\nonumber\\
\end{eqnarray}
Comparison of the upper bounds (\ref{U1}), (\ref{U2}) and
(\ref{U3}) shows that $U_{1}\geq U_{3}$ and $U_{2}\geq U_{3}$ and
therefore, the upper bound (\ref{Ubound}) is equal to $U_{3}$,
i.e., $U(\rho_{AB}^{(1)})=U_{3}(\rho_{AB}^{(1)})$. Because the
bound also represents at homodyne detections of quadratures $x_A$
and $x_B$ on modes $A$ and $B$, the least mutual information over
all Gaussian measurements on mode $E$, the upper bound (\ref{U3})
also gives GIE for all symmetric GLEMS satisfying inequality
(\ref{IcGcondition}),
\begin{equation}\label{GIEsymGLEMS2}
E_{\downarrow}^{G}\left(\rho_{AB}^{(1)}\right)=\ln\left(\frac{a}{\sqrt{a^2-k_p^2}}\right).
\end{equation}
Note, that the RHS of the latter formula is nothing but the mutual
information of the joint distribution of outcomes of measurements
of quadratures $p_A$ and $p_B$ on a symmetric Gaussian state with
CM (\ref{gammasymst}).

In particular, for the CV GHZ state we have
\begin{equation}\label{akxkpCVGHZ}
a=\sqrt{x_{+}x_{-}},\quad
k_{x,p}=\sqrt{\frac{x_{\mp}}{x_{\pm}}}(x_{-}-x_{+})
\end{equation}
which gives after substitution into the RHS of
Eq.~(\ref{GIEsymGLEMS2}) the formula
\begin{eqnarray}\label{GIEGHZ}
E_{\downarrow}^{G}\left(\rho_{AB}^{GHZ}\right)=\ln\left(\frac{x_{-}}{e^{r}\sqrt{x_{+}}}\right).
\end{eqnarray}
The latter expression agrees with the GIE for the CV GHZ state derived in
Ref.~\cite{Mista_15}.

It remains to identify the set of symmetric entangled GLEMS for
which the GIE is given by formula (\ref{GIEsymGLEMS2}). This set
comprise all the states, for which inequality (\ref{IcGcondition})
is satisfied for all CMs $\Gamma_{E}$. In order to find the
states, we can again use the results of Ref.~\cite{Mista_15}.
There, minimum of the left-hand side (LHS) of the inequality for
CV GHZ state was found over all CMs $\Gamma_E$, which was strictly
positive for all values of the squeezing parameter $r$. Hence, the
inequality is satisfied for all CV GHZ states and therefore GIE is
for all the states given by formula (\ref{GIEGHZ}).

Inspired by the approach of Ref.~\cite{Mista_15}, we can also
minimize the LHS of inequality (\ref{IcGcondition}). Like in the
case of quantity (\ref{g}), the LHS depends on variables
$\mathcal{V}_{p},\mathcal{V}_{x}$ and $\varphi$ that we minimize
over in the same way, as the LHS for the CV GHZ state, up to the
already mentioned difference in the form of the parameter $q$.
Therefore, we can minimize the LHS of inequality
(\ref{IcGcondition}) exactly in the same way as the LHS of the
inequality was minimized in Ref.~\cite{Mista_15} for a particular
case of the CV GHZ state. Hence it follows, that also for
symmetric entangled GLEMS both the quantities $\tilde{a}$ and
$\tilde{s}$ attain maximal values of $\tilde{a}_{\rm max}=a$ and
$\tilde{s}_{\rm max}=\sqrt{a^2-k_x^{2}}$ for projection of mode
$E$ onto an infinitely hot thermal state which is equivalent with
dropping of mode $E$. Consequently, the LHS of inequality
(\ref{IcGcondition}) possesses the following lower bound:
\begin{equation}\label{lowerbound}
2+\frac{1}{\tilde{a}}-\tilde{s}\geq
2+\frac{1}{a}-\sqrt{a^2-k_x^{2}}.
\end{equation}
Further, making use of relation (\ref{kxkp}) we see, that for
symmetric GLEMS the necessary and sufficient condition for
entanglement, $1>(a-k_{x})(a-k_{p})$, is equivalent with
inequality $k_{p}>0$. Taking the latter inequality together with
inequality $a\geq1$ and relation (\ref{kxkp}) we then find, that
\begin{equation}\label{lowerbound2}
\sqrt{a^2-k_x^{2}}=\sqrt{2-\frac{2k_{p}(a+k_{p})+1}{(a+k_{p})^2}}<\sqrt{2}.
\end{equation}
Combining now inequalities (\ref{lowerbound}) and
(\ref{lowerbound2}) we arrive at a strictly positive lower bound
for the LHS of inequality (\ref{IcGcondition}) of the form:
\begin{equation}\label{lowerbound3}
2+\frac{1}{\tilde{a}}-\tilde{s}>2-\sqrt{2}\doteq0.586.
\end{equation}
This implies, that inequality (\ref{IcGcondition}) is satisfied
for {\it all} symmetric entangled GLEMS and thus the formula
(\ref{GIEsymGLEMS2}) gives GIE for {\it all} the states.

\subsection{GIE for symmetric squeezed thermal states}\label{subsec_symisotropic}

In this section we calculate GIE for states which have symmetric standard-form CM
(\ref{gammasymst}) and which are isotropic \cite{Botero_03}, i.e., their symplectic
eigenvalues (\ref{nu12sym}) are equal, $\nu_1=\nu_2$. Due to the equality one finds
using Eq.~(\ref{nu12sym}) that $k_x=k_p\equiv k$ and hence the standard-form CM of
the states considered in this section reads as
\begin{eqnarray}\label{gammaisst}
\gamma_{AB}=\left(\begin{array}{cccc}
a & 0 & k & 0\\
0 & a & 0 & -k \\
k & 0 & a & 0 \\
0 & -k & 0 & a \\
\end{array}\right),
\end{eqnarray}
which is nothing but a CM of the well known two-mode symmetric
squeezed thermal states. The CM describes a physical quantum state
if and only if $a^2-k^{2}\geq 1$ and the state is entangled if and
only if $1>a-k$. What is more, CM (\ref{gammaisst}) has a single
two-fold degenerate symplectic eigenvalue
\begin{equation}\label{nuis}
\nu=\sqrt{a^2-k^2}
\end{equation}
and it also holds that the local symplectic transformations
(\ref{SAB}) satisfy $S_{A}=S_{B}^{-1}\equiv Z$, where
\begin{equation}\label{SABis}
Z=\left(\begin{array}{cc}
z^{-1} & 0 \\
0 & z\\
\end{array}\right)
\end{equation}
with $z=\sqrt[4]{\frac{a+k}{a-k}}$.

A particular instance of such states is given by pure two-mode
Gaussian states which satisfy $\nu=1$. As for the pure states GIE
was already calculated in Ref.~\cite{Mista_15}, here we focus on
its evaluation for mixed states for which the following inequality
$\nu>1$ holds. For the latter states one obviously has $R=2$ and
therefore we denote them as $\rho_{AB}^{(2)}$ and the
corresponding standard-form CM as $\gamma_{AB}^{(2)}$ in what
follows. Since $R=2$ for CM $\gamma_{AB}^{(2)}$ the corresponding
states possess a four-mode purification with a purifying system
consisting of two modes $E_{A}$ and $E_{B}$, which we shall for
brevity denote as $E$ in what follows. By performing a two-mode
Gaussian measurement with CM $\Gamma_{E}$ on subsystem $E$, it
follows from Eqs.~(\ref{gammacondS}), (\ref{gammacond0}) and
(\ref{Ssym}), that the purification collapses to a conditional
state $\rho_{AB|E}^{(2)}$ with CM
\begin{eqnarray}\label{gammacondSis}
\gamma_{AB|E}^{(2)}=U_{BS}^{T}(Z^{-1}\oplus
Z)\gamma_{AB|E}^{(0)}[(Z^{-1})^{T}\oplus Z^{T}]U_{BS},\nonumber\\
\end{eqnarray}
where
\begin{eqnarray}\label{gammacond0is}
\gamma_{AB|E}^{(0)}&=&\nu\openone_{4}-(\nu^2-1)\frac{1}{\nu\openone_{4}+\Gamma_{E}^{(T)}},
\end{eqnarray}
where $\openone_{4}$ is the $4\times4$ identity matrix and
$\Gamma_{E}^{(T)}=(\sigma_{z}\oplus\sigma_{z})\Gamma_{E}(\sigma_{z}\oplus\sigma_{z})$
is a CM of the transpose of the state with CM $\Gamma_{E}$.

Like in the previous case we calculate GIE using the upper bound
\begin{equation}\label{Uboundis}
U\left(\rho_{AB}^{(2)}\right)=\mathop{\mbox{inf}}_{\Gamma_{E}}\left[\mathcal{I}_{c}^{G}\left(\rho_{AB|E}^{(2)}\right)\right],
\end{equation}
where
\begin{equation}\label{IcGis}
\mathcal{I}_{c}^{G}\left(\rho_{AB|E}^{(2)}\right)=\mathop{\mbox{sup}}_{\Gamma_{A},\Gamma_{B}}
f(\gamma_{\pi},\Gamma_{A},\Gamma_{B},\Gamma_{E}),
\end{equation}
which is the GCMI of the conditional quantum state
$\rho_{AB|E}^{(2)}$ with CM (\ref{gammacondSis}) \cite{Mista_11}.
Due to the invariance of a determinant with respect to symplectic
transformations we can assume the CM $\gamma_{AB|E}^{(2)}$ of the
latter state to be in the standard form
\begin{eqnarray}\label{gammagenst}
\gamma_{AB|E}^{(2)}=\left(\begin{array}{cccc}
\tilde{a} & 0 & \tilde{c}_x & 0\\
0 & \tilde{a} & 0 & \tilde{c}_p \\
\tilde{c}_x & 0 & \tilde{b} & 0 \\
0 & \tilde{c}_p & 0 & \tilde{b} \\
\end{array}\right),
\end{eqnarray}
where $\tilde{c}_x\geq|\tilde{c}_p|\geq0$. Unfortunately, as in
the present case Eve's measurement generally violates the symmetry
between modes $A$ and $B$, i.e., $\tilde{a}\ne \tilde{b}$, we
cannot use the formula (\ref{IcGsym}) to calculate the GCMI
(\ref{IcGis}). However, inspired by the derivation of formula
(\ref{IcGsym}) performed in Ref.~\cite{Mista_15}, we can derive an
analytic expression for the GCMI even for certain subset of
two-mode Gaussian states with nonsymmetric CMs (\ref{gammagenst}).

We start our derivation by recalling \cite{Mista_11}, that for a
two-mode Gaussian state with the standard-form CM
(\ref{gammagenst}), the GCMI reads as
\begin{equation}\label{IcGgen}
\mathcal{I}_{c}^{G}\left(\rho_{AB|E}^{(2)}\right)=-\frac{1}{2}\ln\left\{\mathop{\mbox{inf}}_{r_{A},r_{B}\geq0}\left[u(r_{A},r_{B})\right]\right\}
\end{equation}
with
\begin{widetext}
\begin{eqnarray}\label{u}
u(r_{A},r_{B})=\left[1-\frac{\tilde{c}_{x}^2}{a_{-}(r_{A})b_{-}(r_{B})}\right]\left[1-\frac{\tilde{c}_{p}^2}{a_{+}(r_{A})b_{+}(r_{B})}\right],
\end{eqnarray}
\end{widetext}
where $a_{\pm}(r_{A})\equiv \tilde{a}+e^{\pm 2r_{A}}$,
$b_{\pm}(r_{B})\equiv \tilde{b}+e^{\pm 2r_{B}}$, and where $r_{A}$
and $r_{B}$ are nonnegative squeezing parameters of Gaussian
measurements with CMs
$\Gamma_{A}=\mbox{diag}(e^{-2r_{A}},e^{2r_{A}})$ and
$\Gamma_{B}=\mbox{diag}(e^{-2r_{B}},e^{2r_{B}})$ on modes $A$ and
$B$. The minimization of the latter function with respect to the
squeezing parameters $r_{A}$ and $r_{B}$ can be done by the
following chain of inequalities:
\begin{widetext}
\begin{eqnarray}\label{chainineq}
\left[1-\frac{\tilde{c}_{x}^2}{a_{-}(r_{A})b_{-}(r_{B})}\right]\left[1-\frac{\tilde{c}_{p}^2}{a_{+}(r_{A})b_{+}(r_{B})}\right]&\geq&
\left[1-\frac{\tilde{c}_{x}^2}{a_{-}(r_{A})b_{-}(r_{B})}\right]\left[1-\frac{\tilde{c}_{x}^2}{a_{+}(r_{A})b_{+}(r_{B})}\right]=1-\frac{\tilde{c}_{x}^2}{\tilde{a}\tilde{b}}
\nonumber\\
&+&\frac{\tilde{c}_{x}^{2}}{\tilde{a}\tilde{b}}
\left\{\frac{p+2\tilde{a}\cosh(2r_{A})+2\tilde{b}\cosh(2r_{B})+2\tilde{a}\tilde{b}\cosh[2(r_{A}-r_{B})]}{\left[1+\tilde{a}^2+2\tilde{a}\cosh(2r_{A})\right]
[1+\tilde{b}^2+2\tilde{b}\cosh(2r_{B})]}\right\}
\nonumber\\
&\geq& 1-\frac{\tilde{c}_{x}^2}{\tilde{a}\tilde{b}},
\end{eqnarray}
\end{widetext}
where
\begin{equation}\label{p}
p\equiv1+\tilde{a}^2+\tilde{b}^2-\tilde{a}\tilde{b}(\tilde{a}\tilde{b}-\tilde{c}_{x}^2).
\end{equation}
Here, the first inequality is a consequence of inequality
$\tilde{c}_{x}\geq |\tilde{c}_{p}|$ and the second inequality
holds if the inequality
\begin{equation}\label{IcGgencond1}
p+2(\tilde{a}+\tilde{b}+\tilde{a}\tilde{b})\geq0
\end{equation}
is satisfied. The latter inequality can be further rewritten into
the compact form
\begin{equation}\label{IcGgencond2}
(\tilde{a}+\tilde{b}+1)^2\geq
\tilde{a}\tilde{b}(\tilde{a}\tilde{b}-\tilde{c}_{x}^2)
\end{equation}
which is equivalent with inequality
\begin{equation}\label{IcGgencond}
G\equiv\sqrt{\frac{\tilde{a}}{\tilde{b}}}+\sqrt{\frac{\tilde{b}}{\tilde{a}}}+\frac{1}{\sqrt{\tilde{a}\tilde{b}}}-\sqrt{\tilde{a}\tilde{b}-\tilde{c}_{x}^2}\geq0.
\end{equation}

Importantly, the lower bound
$1-\tilde{c}_{x}^2/(\tilde{a}\tilde{b})$ in inequalities
(\ref{chainineq}) is tight because it can be achieved in the limit
$r_{A}\rightarrow+\infty$ and $r_{B}\rightarrow+\infty$, which
corresponds to the double homodyne detection of quadratures
$x_{A}$ and $x_{B}$ on modes $A$ and $B$. We have thus arrived to
the finding that for all states with CM (\ref{gammagenst}) for
which the parameters $\tilde{a},\tilde{b}$ and $\tilde{c}_{x}$
satisfy inequality (\ref{IcGgencond}), the optimal measurement in
GCMI (\ref{IcGis}) is double homodyne detection of
$x$-quadratures. Hence, one gets for this class of states
\begin{equation}\label{IcGgen2}
\mathcal{I}_{c}^{G}\left(\rho_{AB|E}^{(2)}\right)=\frac{1}{2}\ln\frac{\tilde{a}\tilde{b}}{\tilde{a}\tilde{b}-\tilde{c}_{x}^2}.
\end{equation}

Note, that in Ref.~\cite{Mista_11} an analytical formula for GCMI
was derived for two-mode squeezed thermal states with CM
(\ref{gammagenst}), where $\tilde{c}_{p}=-\tilde{c}_{x}$. Later,
in Ref.~\cite{Mista_15} the GCMI was calculated for two-mode
symmetric states with $\tilde{a}=\tilde{b}$, which satisfy
inequality (\ref{IcGcondition}). Our formula (\ref{IcGgen2}) and
condition (\ref{IcGgencond}) thus represent a generalization of
the latter result to generic nonsymmetric two-mode Gaussian
states.

Obviously, formula (\ref{IcGgen2}) is useful for evaluation of the
upper bound (\ref{Uboundis}) only for those considered states, for
which inequality (\ref{IcGgencond}) is satisfied for any CM
$\Gamma_{E}$. In order to find some of these states we first
consider the following chain of inequalities:
\begin{equation}\label{chainineq2}
G\geq\sqrt{\frac{\tilde{a}}{\tilde{b}}}+\sqrt{\frac{\tilde{b}}{\tilde{a}}}+\frac{1}{\sqrt{\tilde{a}\tilde{b}}}-\sqrt{\tilde{a}\tilde{b}}\geq
2-\left(\sqrt{\tilde{a}\tilde{b}}-\frac{1}{\sqrt{\tilde{a}\tilde{b}}}\right)
\end{equation}
where the second inequality is a consequence of the inequality
\begin{equation}\label{aritgeomineq}
\sqrt{\frac{\tilde{a}}{\tilde{b}}}+\sqrt{\frac{\tilde{b}}{\tilde{a}}}=\frac{\tilde{a}+\tilde{b}}{\sqrt{\tilde{a}\tilde{b}}}\geq2,
\end{equation}
which follows from the inequality of arithmetic and geometric
means. It is further convenient to define a nonnegative squeezing
parameter $s$ by the formula
$e^{s}=\sqrt{\tilde{a}\tilde{b}}\geq1$ which allows us to express
the RHS of inequality~(\ref{chainineq2}) as
\begin{equation}\label{sinhs}
2-\left(\sqrt{\tilde{a}\tilde{b}}-\frac{1}{\sqrt{\tilde{a}\tilde{b}}}\right)=2[1-\sinh(s)].
\end{equation}
Hence, if $1\geq\sinh(s)$, the RHS of the chain of inequalities
(\ref{chainineq2}) is nonnegative and therefore inequality
(\ref{IcGgencond}) is satisfied. The monotonicity of the inverse
hyperbolic sine function further allows us to rewrite the latter
inequality into an equivalent inequality
$0.881\doteq\sinh^{-1}(1)\geq s=\ln\sqrt{\tilde{a}\tilde{b}}$
which can be finally rewritten due to the monotonicity of the
exponential function as
\begin{equation}\label{sqrtabbound}
\sqrt{\tilde{a}\tilde{b}}\leq2.41,
\end{equation}
where we have used that $e^{0.881}\doteq2.41$. Consequently, if
for a conditional state $\rho_{AB|E}^{(2)}$ the parameters
$\tilde{a}$ and $\tilde{b}$ of CM (\ref{gammagenst}) satisfy
inequality (\ref{sqrtabbound}) for all CMs $\Gamma_{E}$, the GCMI
(\ref{IcGis}) is given by formula (\ref{IcGgen2}) for all CMs
$\Gamma_{E}$ and the upper bound (\ref{Uboundis}) can be derived
by minimization of the RHS of Eq.~(\ref{IcGgen2}) over all CMs
$\Gamma_{E}$.

The expression $\sqrt{\tilde{a}\tilde{b}}$ on the LHS of
inequality (\ref{sqrtabbound}) depends in a complicated way on CM
$\Gamma_E$ and hence it does not allow us to identify for which of
the considered states the inequality will be satisfied for any
$\Gamma_{E}$. However, if we find an upper bound on the expression
which would depend only on parameters of CM $\gamma_{AB}^{(2)}$
but which would be independent of CM $\Gamma_{E}$, then for all
the states for which the upper bound is less than or equal to
$2.41$ the inequality (\ref{sqrtabbound}) will be satisfied for
any $\Gamma_{E}$. What is more, provided that the bound is simple,
the obtained inequality would give us a simple characterization of
a subset of the considered set of states for which inequality
(\ref{IcGgencond}) is satisfied for any CM $\Gamma_{E}$ as
required.

Interestingly, such an upper bound can be really found which
is on the top of that tight. To show this note first, that for the
parameters $\tilde{a}$ and $\tilde{b}$ appearing on the LHS of
inequality (\ref{sqrtabbound}) it holds that
$\tilde{a}=\sqrt{\mbox{det}\tilde{A}}$ and
$\tilde{b}=\sqrt{\mbox{det}\tilde{B}}$, where $\tilde{A}$ and
$\tilde{B}$ are diagonal $2\times2$ blocks of CM
(\ref{gammacondSis}) expressed in the block form with respect to
$A|B$ splitting,
\begin{eqnarray}\label{gammasblock2}
\gamma_{AB|E}^{(2)}=\left(\begin{array}{cc}
\tilde{A} & \tilde{C} \\
\tilde{C}^{T} & \tilde{B} \\
\end{array}\right).
\end{eqnarray}
By expressing also CM (\ref{gammacond0is}) in the block form with
respect to the same splitting,
\begin{eqnarray}\label{gammasblock3}
\gamma_{AB|E}^{(0)}=\left(\begin{array}{cc}
\mathcal{A} & \mathcal{C} \\
\mathcal{C}^{T} & \mathcal{B} \\
\end{array}\right)
\end{eqnarray}
and substituting the expression into the RHS of
Eq.~(\ref{gammacondSis}), we further get using
Eq.~(\ref{Ubeamsplitter}) the matrices $\tilde{A}$ and $\tilde{B}$
in the form
\begin{eqnarray}\label{tildeAB}
\tilde{A}=\frac{1}{2}(X-Y),\quad\tilde{B}=\frac{1}{2}(X+Y),
\end{eqnarray}
where
\begin{eqnarray}\label{XY}
X&=&Z^{-1}\mathcal{A}(Z^{-1})^{T}+Z\mathcal{B}Z^{T},\nonumber\\
Y&=&Z^{-1}\mathcal{C}Z^{T}+Z^{T}\mathcal{C}^{T}(Z^{-1})^{T}.
\end{eqnarray}
Now, we make use of the fact that the expression
$\sqrt{\tilde{a}\tilde{b}}$ is bounded from above as
\begin{equation}\label{chainineq3}
\frac{\sqrt{\mbox{det}(\tilde{A}+\tilde{B})}}{2}\geq\frac{\sqrt{\mbox{det}\tilde{A}}+\sqrt{\mbox{det}\tilde{B}}}{2}=\frac{\tilde{a}+\tilde{b}}{2}\geq\sqrt{\tilde{a}\tilde{b}},
\end{equation}
where the first inequality is a consequence of the Minkowski
determinant theorem \cite{Cover_06}
\begin{equation}\label{Minkowski}
\left[\mbox{det}(P+Q)\right]^{\frac{1}{n}}\geq\left(\mbox{det}P\right)^{\frac{1}{n}}+\left(\mbox{det}Q\right)^{\frac{1}{n}},
\end{equation}
which is valid for any symmetric positive semidefinite $n\times n$
matrices $P$ and $Q$, whereas the second inequality is the
inequality of arithmetic and geometric means. Further, inserting
for the matrices $\tilde{A}$ and $\tilde{B}$ from
Eqs.~(\ref{tildeAB}) and (\ref{XY}) into the LHS of inequalities
(\ref{chainineq3}) and using the formula for determinant of a sum
of two $2\times 2$ symmetric matrices $P$ and $Q$
\cite{Fiurasek_07},
\begin{eqnarray}\label{detsum}
\mbox{det}(P+Q)=\mbox{det}P+\mbox{det}Q+\mbox{Tr}(PJQJ^{T}),
\end{eqnarray}
one gets
\begin{eqnarray}\label{detAplusB}
\mbox{det}(\tilde{A}+\tilde{B})&=&\mbox{Tr}\left[Z^{-1}\mathcal{A}(Z^{-1})^{T}JZ\mathcal{B}Z^{T}J^{T}\right]\nonumber\\
&&+\mbox{det}\mathcal{A}+\mbox{det}\mathcal{B}.
\end{eqnarray}
If we now express the CM $\Gamma_{E}^{(T)}$ appearing in
Eq.~(\ref{gammacond0is}) in the following block form with respect
to $A|B$ splitting,
\begin{eqnarray}\label{GammaET}
\Gamma_{E}^{(T)}=\left(\begin{array}{cc}
\alpha & \gamma \\
\gamma^{T} & \beta \\
\end{array}\right),
\end{eqnarray}
and we use the blockwise inversion \cite{Horn_85}
\begin{widetext}
\begin{eqnarray}\label{blockwise}
\left(\begin{array}{cc}
A & C\\
C^{T} & B\\
\end{array}\right)^{-1}=\left(\begin{array}{cc}
\left(A-CB^{-1}C^{T}\right)^{-1} & A^{-1}C\left(C^{T}A^{-1}C-B\right)^{-1}\\
\left(C^{T}A^{-1}C-B\right)^{-1}C^{T}A^{-1} & \left(B-C^{T}A^{-1}C\right)^{-1}\\
\end{array}\right),
\end{eqnarray}
\end{widetext}
we can express the local $2\times2$ CMs $\mathcal{A}$ and
$\mathcal{B}$ of modes $A$ and $B$ as
\begin{eqnarray}\label{calAB}
\mathcal{A}&=&\nu\openone-(\nu^{2}-1)\left(\nu\openone+\Delta_{\mathcal{A}}\right)^{-1},\nonumber\\
\mathcal{B}&=&\nu\openone-(\nu^{2}-1)\left(\nu\openone+\Delta_{\mathcal{B}}\right)^{-1},
\end{eqnarray}
where
\begin{eqnarray}\label{DeltaAB}
\Delta_{\mathcal{A}}&=&\alpha-\gamma(\beta+\nu\openone)^{-1}\gamma^{T},\nonumber\\
\Delta_{\mathcal{B}}&=&\beta-\gamma^{T}(\alpha+\nu\openone)^{-1}\gamma.
\end{eqnarray}
The matrices are nothing but CMs of conditional quantum states
obtained by a measurement with CM $\nu\openone$ of a local
subsystem of a system in a quantum state with CM (\ref{GammaET}).
The matrices (\ref{DeltaAB}) thus possess nonnegative eigenvalues
and if we denote them as
$\mbox{eig}(\Delta_{\mathcal{A}})=\{\mu_{1},\mu_{2}\}$, where
$\mu_{1}\geq\mu_{2}$, and
$\mbox{eig}(\Delta_{\mathcal{B}})=\{\varepsilon_{1},\varepsilon_{2}\}$,
where $\varepsilon_{1}\geq\varepsilon_{2}$, we find that the
eigenvalues of CMs (\ref{calAB}) read as
\begin{eqnarray}\label{eigcalAB}
\mbox{eig}\left(\mathcal{A}\right)&=&\left\{a_{j}=\frac{\nu\mu_{j}+1}{\nu+\mu_{j}},\quad
j=1,2\right\},\nonumber\\
\mbox{eig}\left(\mathcal{B}\right)&=&\left\{b_{j}=\frac{\nu\varepsilon_{j}+1}{\nu+\varepsilon_{j}},\quad
j=1,2\right\}
\end{eqnarray}
and they satisfy inequalities $a_{1}\geq a_{2}$ and $b_{1}\geq
b_{2}$ which follow from inequalities $\mu_{1}\geq\mu_{2}$ and
$\varepsilon_{1}\geq\varepsilon_{2}$, respectively. Consequently,
the determinants appearing on the RHS of Eq.~(\ref{detAplusB})
possess the following upper bounds:
\begin{eqnarray}\label{detABbound}
\mbox{det}\mathcal{A}&=&a_{1}a_{2}=\nu^{2}\left(\frac{\mu_{1}+\frac{1}{\nu}}{\mu_{1}+\nu}\right)\left(\frac{\mu_{2}+\frac{1}{\nu}}{\mu_{2}+\nu}\right)\leq\nu^{2},\nonumber\\
\mbox{det}\mathcal{B}&=&b_{1}b_{2}=\nu^{2}\left(\frac{\varepsilon_{1}+\frac{1}{\nu}}{\varepsilon_{1}+\nu}\right)
\left(\frac{\varepsilon_{2}+\frac{1}{\nu}}{\varepsilon_{2}+\nu}\right)\leq\nu^{2},\nonumber\\
\end{eqnarray}
as follows from inequality $\nu\geq1$ which holds for any
symplectic eigenvalue of a physical CM \cite{Simon_94}.

The derivation of the sought upper bound requires finally to find
an upper bound on the trace on the RHS of Eq.~(\ref{detAplusB}).
For this purpose we first use the Cauchy-Schwarz inequality for
real symmetric positive definite matrices $P$ and $Q$
\cite{Bellman_80},
\begin{equation}\label{CS}
\mbox{Tr}(PQ)\leq\sqrt{\mbox{Tr}(P^2)}\sqrt{\mbox{Tr}(Q^2)},
\end{equation}
which gives
\begin{widetext}
\begin{eqnarray}\label{CS2}
\mbox{Tr}\left[Z^{-1}\mathcal{A}(Z^{-1})^{T}JZ\mathcal{B}Z^{T}J^{T}\right]\leq\sqrt{\mbox{Tr}\left\{\left[Z^{-1}\mathcal{A}(Z^{-1})^{T}\right]^2\right\}}
\sqrt{\mbox{Tr}\left[\left(Z\mathcal{B}Z^{T}\right)^2\right]},
\end{eqnarray}
\end{widetext}
where we have used the cyclic property of a trace and identity
$JJ^{T}=\openone$. In the next step, we express CMs $\mathcal{A}$
and $\mathcal{B}$ as
$\mathcal{A}=P(\varphi_{A})\mbox{diag}(a_{1},a_{2})P^{T}(\varphi_{A})$
and
$\mathcal{B}=P(\varphi_{B})\mbox{diag}(b_{1},b_{2})P^{T}(\varphi_{B})$,
where $P(\varphi_{A})$ and $P(\varphi_{B})$ are suitable phase
shifts (\ref{UV}). Making use of the latter decompositions of CMs
$\mathcal{A}$ and $\mathcal{B}$ together with an explicit form of
squeezing transformation $Z$, Eq.~(\ref{SABis}), we can further
bring the CMs in the arguments of the traces on the RHS of
inequality (\ref{CS2}) to the form
\begin{widetext}
\begin{eqnarray}\label{ZcalAB}
Z^{-1}\mathcal{A}(Z^{-1})^{T}&=&\left(\begin{array}{cc}
z^2[a_{+}+a_{-}\cos(2\varphi_{A})] & a_{-}\sin(2\varphi_{A}) \\
a_{-}\sin(2\varphi_{A}) &  \frac{1}{z^2}[a_{+}-a_{-}\cos(2\varphi_{A})] \\
\end{array}\right),\nonumber\\
Z\mathcal{B}Z^{T}&=&\left(\begin{array}{cc}
\frac{1}{z^2}[b_{+}+b_{-}\cos(2\varphi_{B})] & b_{-}\sin(2\varphi_{B}) \\
b_{-}\sin(2\varphi_{B}) &  z^2[b_{+}-b_{-}\cos(2\varphi_{B})] \\
\end{array}\right),
\end{eqnarray}
\end{widetext}
where $a_{\pm}=(a_{1}\pm a_{2})/2$ and $b_{\pm}=(b_{1}\pm
b_{2})/2$. Hence, we arrive after some algebra at the following
upper bounds on the traces:
\begin{widetext}
\begin{eqnarray}\label{Trupperboundhelp}
\mbox{Tr}\left\{\left[Z^{-1}\mathcal{A}(Z^{-1})^{T}\right]^2\right\}&=&a_{-}^{2}\cos^{2}(2\varphi_{A})\left(z^{2}-\frac{1}{z^{2}}\right)^{2}
+2a_{+}a_{-}\cos(2\varphi_{A})\left(z^{4}-\frac{1}{z^{4}}\right)+a_{+}^{2}\left(z^{4}+\frac{1}{z^{4}}\right)+2a_{-}^{2}\nonumber\\
&\leq&a_{-}^{2}\left(z^{2}-\frac{1}{z^{2}}\right)^{2}+2a_{+}a_{-}\left(z^{4}-\frac{1}{z^{4}}\right)+a_{+}^{2}\left(z^{4}+\frac{1}{z^{4}}\right)+2a_{-}^{2}\nonumber\\
&=&a_{1}^{2}z^{4}+\frac{a_{2}^{2}}{z^{4}}\leq\nu^{2}\left(z^{4}+\frac{1}{z^{4}}\right)=2(a^{2}+k^{2}),\nonumber\\
\mbox{Tr}\left[\left(Z\mathcal{B}Z^{T}\right)^2\right]&=&b_{-}^{2}\cos^{2}(2\varphi_{B})\left(z^{2}-\frac{1}{z^{2}}\right)^{2}
-2b_{+}b_{-}\cos(2\varphi_{B})\left(z^{4}-\frac{1}{z^{4}}\right)+b_{+}^{2}\left(z^{4}+\frac{1}{z^{4}}\right)+2b_{-}^{2}\nonumber\\
&\leq&b_{-}^{2}\left(z^{2}-\frac{1}{z^{2}}\right)^{2}+2b_{+}b_{-}\left(z^{4}-\frac{1}{z^{4}}\right)+b_{+}^{2}\left(z^{4}+\frac{1}{z^{4}}\right)+2b_{-}^{2}\nonumber\\
&=&b_{1}^{2}z^{4}+\frac{b_{2}^{2}}{z^{4}}\leq\nu^{2}\left(z^{4}+\frac{1}{z^{4}}\right)=2(a^{2}+k^{2}).\nonumber\\
\end{eqnarray}
\end{widetext}
Here, the second inequalities follow from the fact that
$a_{1,2}\leq\nu$ and $b_{1,2}\leq\nu$, whereas to derive the last
equalities we have used the expression of the symplectic
eigenvalue $\nu$ given in Eq.~(\ref{nuis}) and the squeezing
parameter $z$ given below Eq.~(\ref{SABis}). If we now use the
latter two upper bounds on the RHS of inequality (\ref{CS2}) we
find that
\begin{eqnarray}\label{Trupperbound}
\mbox{Tr}\left[Z^{-1}\mathcal{A}(Z^{-1})^{T}JZ\mathcal{B}Z^{T}J^{T}\right]\leq2(a^{2}+k^{2}).
\end{eqnarray}
Finally, combining inequalities (\ref{chainineq3}),
(\ref{detABbound}) and (\ref{Trupperbound}) with equality
(\ref{detAplusB}), we find that
\begin{equation}\label{Finalupperbound}
\sqrt{\tilde{a}\tilde{b}}\leq a,
\end{equation}
where $a$ is the parameter of the standard form (\ref{gammaisst})
of CM $\gamma_{AB}^{(2)}$. Note, that
the latter upper bound is intuitive and tight because it is
reached if Eve drops her two modes (which is equivalent with
projection of the modes onto infinitely hot thermal states).

The derived upper bound (\ref{Finalupperbound}) now allows us to
identify a set of two-mode symmetric squeezed thermal states for
which the GCMI (\ref{IcGis}) is given by formula (\ref{IcGgen2})
for any CM $\Gamma_{E}$. Concretely, if for the parameter $a$ of
CM $\gamma_{AB}^{(2)}$ it holds that
\begin{equation}\label{Finalupperbound2}
a\leq 2.41,
\end{equation}
then inequality (\ref{sqrtabbound}) and hence also inequality
(\ref{IcGgencond}) is satisfied for any CM $\Gamma_{E}$.
Consequently, for all two-mode symmetric squeezed thermal states
for which inequality (\ref{Finalupperbound2}) is fulfilled GCMI
(\ref{IcGis}) is given by formula (\ref{IcGgen2}) for any CM
$\Gamma_{E}$ as we set out to prove.

Before going further needles to say that CM (\ref{GammaET}) we
have been working with above is a CM of a Gaussian measurement.
This means that it can be not only a CM of a physical quantum
state which is invertible, but also a CM of a nonphysical quantum
state. This includes, for instance, infinitely squeezed state
which describes homodyne detection or a thermal state with
infinite mean number of thermal photons, which corresponds to
dropping of the measured system. Since nonphysical states can be
obtained as a limit of physical states we usually treat them such
that first we perform the calculation with the corresponding
physical CM and then we take the respective limit at the end of
our calculation. For this reason we used formulas for invertible
matrices in previous derivation of the upper bound
(\ref{Finalupperbound2}) despite the presence of the pseudoinverse
in Eq.~(\ref{gammacond0is}).

We now move to the final step of evaluation of the upper bound
$U(\rho_{AB}^{(2)})$, which is a minimization on the RHS of
Eq.~(\ref{Uboundis}), where GCMI
$\mathcal{I}_{c}^{G}(\rho_{AB|E}^{(2)})$ is given by formula
(\ref{IcGgen2}). At the outset, we express the blocks
$\gamma_{AB}$ and $\gamma_{ABE}$ of CM (\ref{gammapi}) as
\begin{eqnarray}\label{gammaABgammaABE}
\gamma_{AB}=\left(\begin{array}{cc}
\gamma_{A} & \omega_{AB}  \\
\omega_{AB}^{T} & \gamma_{B} \\
\end{array}\right),\quad \gamma_{ABE}=\left(\begin{array}{c}
\gamma_{AE}\\
\gamma_{BE}\\
\end{array}\right).
\end{eqnarray}
Next, we apply to the RHS of Eq.~(\ref{xSigma}) the determinant
formula \cite{Horn_85}:
\begin{equation}\label{det}
\mbox{det}(M)=\mbox{det}(\mathfrak{D})\mbox{det}(\mathfrak{A}-\mathfrak{B}\mathfrak{D}^{-1}\mathfrak{C}),
\end{equation}
which is valid for any $(n+m)\times(n+m)$ matrix
\begin{eqnarray}\label{M}
M=\left(\begin{array}{cc}
\mathfrak{A} & \mathfrak{B}\\
\mathfrak{C} & \mathfrak{D}\\
\end{array}\right),
\end{eqnarray}
where $\mathfrak{A}$, $\mathfrak{B}$ and $\mathfrak{C}$ are
respectively $n\times n$, $n\times m$ and $m\times n$ matrices and
$\mathfrak{D}$ is an $m\times m$ invertible matrix. This allows us
to express Eq.~(\ref{f}) as
\begin{eqnarray}\label{fgen}
f\left(\gamma_{\pi},\Gamma_{A},\Gamma_{B},\Gamma_{E}\right)=I(A;B)+K(E|A;B),
\end{eqnarray}
where
\begin{eqnarray}\label{IAB}
I(A;B)=\frac{1}{2}\ln\left[\frac{\mbox{det}(\Gamma_{A}+\gamma_{A})\mbox{det}(\Gamma_{B}+\gamma_{B})}{\mbox{det}\left(\Gamma_{A}\oplus\Gamma_{B}+\gamma_{AB}\right)}\right]
\end{eqnarray}
and
\begin{equation}\label{KEABhelp}
K(E|A;B)=\frac{1}{2}\ln \mathscr{K}
\end{equation}
with
\begin{eqnarray}\label{K1}
\mathscr{K}=\frac{\mbox{det}(\Gamma_{E}+X_{A})\mbox{det}(\Gamma_{E}+X_{B})}{\mbox{det}(\Gamma_{E}+X_{AB})\mbox{det}(\Gamma_{E}+\gamma_{E})}.
\end{eqnarray}
Here
\begin{eqnarray}\label{Xj}
X_{j}=\gamma_{E}-\gamma_{jE}^{T}(\Gamma_{j}+\gamma_{j})^{-1}\gamma_{jE},\quad
j=A,B,
\end{eqnarray}
and
\begin{eqnarray}\label{XAB}
X_{AB}=\gamma_{E}-\gamma_{ABE}^{T}(\Gamma_{A}\oplus\Gamma_{B}+\gamma_{AB})^{-1}
\gamma_{ABE}.
\end{eqnarray}
The advantage of the alternative expression (\ref{fgen}) for the
optimized function $f$ is that it depends on CM $\Gamma_{E}$ we
minimize over in a simple way via quantity (\ref{K1}). From our
previous results we know that the GCMI to be minimized over all
CMs $\Gamma_{E}$, which is given in Eq.~(\ref{IcGgen2}), is
obtained from Eq.~(\ref{fgen}) if Alice and Bob perform homodyne
detection of quadratures $x_{A}$ and $x_{B}$ on their modes. The
latter measurement is described by CM
$\Gamma_{A}\oplus\Gamma_{B}\equiv\Gamma_{x_{A}}^{t_{A}}\oplus\Gamma_{x_{B}}^{t_{B}}$,
where
$\Gamma_{x_{j}}^{t_{j}}\equiv\mbox{diag}(e^{-2t_{j}},e^{2t_{j}})$,
$j=A,B$, in the limit for $t_{A}\rightarrow+\infty$ and
$t_{B}\rightarrow+\infty$. Thus, setting
$\Gamma_{A}=\Gamma_{x_{A}}^{t_{A}}$ and
$\Gamma_{B}=\Gamma_{x_{B}}^{t_{B}}$ in Eq.~(\ref{fgen}) and
carrying out the limits one finds that
\begin{eqnarray}\label{IcGis2}
\mathcal{I}_{c}^{G}\left(\rho_{AB|E}^{(2)}\right)=I_{\rm h}(A;B)+K_{\rm h}(E|A;B),
\end{eqnarray}
where
\begin{equation}\label{IABis}
I_{\rm h}(A;B)=\frac{1}{2}\ln\frac{a^2}{a^2-k^2}
\end{equation}
and
\begin{equation}\label{KEAB}
K_{\rm h}(E|A;B)=\frac{1}{2}\ln \mathscr{K}_{\rm h},
\end{equation}
where $\mathscr{K}_{\rm h}$ is given by the RHS of Eq.~(\ref{K1})
with
\begin{eqnarray}\label{XAXBhom}
X_{A,B}=\left(\begin{array}{cccc}
\nu-\frac{\nu^{2}-1}{2a}z^2 & 0 & \pm\frac{\nu^{2}-1}{2a} & 0\\
0 & \nu & 0 & 0 \\
\pm\frac{\nu^{2}-1}{2a} & 0 & \nu-\frac{\nu^{2}-1}{2a}\frac{1}{z^2} & 0 \\
0 & 0 & 0 & \nu \\
\end{array}\right)
\end{eqnarray}
and $X_{AB}$ being the following diagonal matrix:
\begin{equation}\label{XABhom}
X_{AB}=\mbox{diag}\left(\frac{1}{\nu},\nu,\frac{1}{\nu},\nu\right).
\end{equation}
Combining Eqs.~(\ref{Uboundis}), (\ref{IcGis2}), (\ref{IABis}) and
(\ref{KEAB}) and using the monotonicity of the logarithmic
function one then finds that the sought upper bound is equal to
\begin{equation}\label{Uboundis2}
U\left(\rho_{AB}^{(2)}\right)=\frac{1}{2}\ln\frac{a^2}{a^2-k^2}+\frac{1}{2}\ln\mathscr{K}_{\rm
min},
\end{equation}
where
\begin{equation}\label{Kmin}
\mathscr{K}_{\rm
min}\equiv\mathop{\mbox{inf}}_{\Gamma_{E}}\mathscr{K}_{\rm h}.
\end{equation}

Before carrying out the latter minimization we first simplify to a
maximum extent the function $\mathscr{K}_{\rm h}$ that is to be
minimized. So far, we have used the form of CMs which has in the
$n$-mode case elements defined as
$\gamma_{jk}=\langle\Delta\xi_{j}\Delta\xi_{k}+\Delta\xi_{k}\Delta\xi_{j}\rangle$,
$j,k=1,2,\ldots,2n$, with
$\Delta\xi_{j}=\xi_{j}-\langle\xi_{j}\rangle$, where
$\xi=(x_{1},p_{1},\ldots,x_{n},p_{n})^{T}$ is a column vector
of quadrature operators of the $n$ modes. As the matrices
(\ref{XAXBhom}) contain correlations only between $x$-quadratures,
it is convenient to reorder them and work with CMs with elements
defined as
$\gamma_{jk}'=\langle\Delta\tau_{j}\Delta\tau_{k}+\Delta\tau_{k}\Delta\tau_{j}\rangle$,
where
$\tau=(x_{1},x_{2},\ldots,x_{n},p_{1},p_{2},\ldots,p_{n})^{T}$.
In the present case of two-mode CMs the CMs $\gamma$ and $\gamma'$
are related as $\gamma'=\Lambda\gamma\Lambda^{T}$ with $\Lambda$
being a matrix
\begin{eqnarray}\label{Lambda}
\Lambda=\left(\begin{array}{cccc}
1 & 0 & 0 & 0\\
0 & 0 & 1 & 0 \\
0 & 1 & 0 & 0 \\
0 & 0 & 0 & 1\\
\end{array}\right).
\end{eqnarray}
Hence, after the reordering CM (\ref{XABhom}) reads as
follows:
\begin{eqnarray}\label{XABprimed}
X_{AB}'=\left(\frac{1}{\nu}\openone\right)\oplus(\nu\openone),
\end{eqnarray}
whereas CMs (\ref{XAXBhom}) change to
\begin{eqnarray}\label{XAXBprimed}
X_{A}'&=&\gamma_{E}-|\chi_{A}\rangle\langle\chi_{A}|,\nonumber\\
X_{B}'&=&X_{AB}'+|\chi_{B}\rangle\langle\chi_{B}|,
\end{eqnarray}
where $\gamma_{E}=\nu\openone_{4}$,
\begin{eqnarray}\label{chiAB}
|\chi_{A,B}\rangle=\sqrt{2\sinh(v)}\left(\begin{array}{c}
|e_{A,B}\rangle\\
\mathbb{O}_{2\times1}\\
\end{array}\right)
\end{eqnarray}
with $\sinh(v)\equiv(\nu-1/\nu)/2$, $v>0$, and
\begin{eqnarray}\label{eAB}
|e_{A}\rangle=\sigma_{x}|e_{B}\rangle=\frac{1}{\sqrt{z^4+1}}\left(\begin{array}{c}
z^2 \\
-1 \\
\end{array}\right),
\end{eqnarray}
where $\sigma_{x}$ is the Pauli-$x$ matrix. Since the matrix
$\Lambda$, Eq.~(\ref{Lambda}), is orthogonal and a determinant is
invariant with respect to orthogonal transformations we can
express the quantity $\mathscr{K}_{\rm h}$ in the new ordering as
\begin{eqnarray}\label{K2}
\mathscr{K}_{\rm h}&=&\frac{\mbox{det}(\Gamma_{E}'+\gamma_{E}-|\chi_{A}\rangle\langle\chi_{A}|)}{\mbox{det}(\Gamma_{E}'+\gamma_{E})}\nonumber\\
&&\times\frac{\mbox{det}(\Gamma_{E}'+X_{AB}'+|\chi_{B}\rangle\langle\chi_{B}|)}{\mbox{det}(\Gamma_{E}'+X_{AB}')},
\end{eqnarray}
where $\Gamma_{E}'=\Lambda\Gamma_{E}\Lambda^{T}$. The advantage of
the expression on the RHS of equation (\ref{K2}) is that the
matrix in the argument of the first (second) determinant in the
nominator is a difference (sum) of a matrix in the argument of the
first (second) determinant in the denominator and a rank-1 matrix
$|\chi_{A}\rangle\langle\chi_{A}|$
($|\chi_{B}\rangle\langle\chi_{B}|$). This allows us to
considerably simplify quantity $\mathscr{K}_{\rm h}$. Namely, by
applying the determinant formula (\ref{det}) on the following
unitarily equivalent matrices
\begin{eqnarray}\label{M1M2}
M_{1}=\left(\begin{array}{cc}
\mathcal{X} & -|c\rangle\\
\langle r| & 1\\
\end{array}\right)\,\,\mbox{and}\,\, M_{2}=\left(\begin{array}{cc}
1 & \langle r|\\
-|c\rangle & \mathcal{X}\\
\end{array}\right),
\end{eqnarray}
where $\mathcal{X}$ is an invertible matrix and $|c\rangle$ and
$\langle r|$ is a column and row vector, respectively, we arrive
at the formula \cite{Henderson_81}
\begin{equation}\label{matdetlem}
\mbox{det}(\mathcal{X}+|c\rangle\langle r|)=\left(1+\langle
r|\mathcal{X}^{-1}|c\rangle\right)\mbox{det}\mathcal{X}.
\end{equation}
By applying now the latter formula on the determinants in the
nominator on the RHS of Eq.~(\ref{K2}) we find that the quantity
(\ref{K2}) simplifies to
\begin{eqnarray}\label{K3}
\mathscr{K}_{\rm h}&=&[1-\langle\chi_{A}|(\Gamma_{E}'+\gamma_{E})^{-1}|\chi_{A}\rangle]\nonumber\\
&&\times[1+\langle\chi_{B}|(\Gamma_{E}'+X_{AB}')^{-1}|\chi_{B}\rangle].
\end{eqnarray}
If we now use in the latter formula equality
$\gamma_{E}=\nu\openone_{4}$ together with Eq.~(\ref{XABprimed}),
we express CM $\Gamma_{E}'$ in the block form with respect to
$x|p$ splitting,
\begin{eqnarray}\label{GammaEprimed}
\Gamma_{E}'=\left(\begin{array}{cc}
\tilde{\alpha} & \tilde{\gamma} \\
\tilde{\gamma}^{T} & \tilde{\beta} \\
\end{array}\right),
\end{eqnarray}
we take into account the formula (\ref{chiAB}) and we use the
blockwise inversion (\ref{blockwise}), we find after some algebra
the quantity (\ref{K3}) to be
\begin{eqnarray}\label{K4}
\mathscr{K}_{\rm h}&=& \left[1+2\sinh(v)\langle
e_{A}|\sigma_{x}\left(\frac{1}{\nu}\openone+\mathcal{Q}\right)^{-1}\sigma_{x}|e_{A}\rangle\right]\nonumber\\
&&\times\left[1-2\sinh(v)\langle
e_{A}|(\nu\openone+\mathcal{Q})^{-1}|e_{A}\rangle\right],
\end{eqnarray}
where
\begin{eqnarray}\label{mathcalQ}
\mathcal{Q}\equiv\tilde{\alpha}-\tilde{\gamma}(\tilde{\beta}+\nu\openone)^{-1}\tilde{\gamma}^{T}.
\end{eqnarray}

Our task is now to minimize the quantity (\ref{K4}) over all CMs
(\ref{GammaEprimed}). Note first, that the RHS of formula
(\ref{K4}) does not depend on the entire CM $\Gamma_{E}'$ but only
on the $2\times 2$ matrix (\ref{mathcalQ}) which consists of
$2\times 2$ submatrices $\tilde{\alpha}, \tilde{\beta}$ and
$\tilde{\gamma}$ of the CM. Since $\Gamma_{E}'$ is a CM, the
submatrix $\tilde{\alpha}$ has the meaning of a classical
correlation matrix of a bivariate Gaussian distribution of
outcomes of homodyne detections of a pair of $x$-quadratures and
therefore it can be an arbitrary positive semidefinite matrix.
Similarly, the off-diagonal submatrix $\tilde{\gamma}$ describes
correlations between the $x$- and $p$-quadratures. Consider now an
admissible pure-state CM given by the following direct sum
$\Gamma_{E}'=\tilde{\alpha}\oplus(\tilde{\alpha})^{-1}$, where
$\tilde{\alpha}$ is an arbitrary positive definite matrix, i.e.,
$\tilde{\alpha}=P(\varphi)\mbox{diag}(e^{-2t_{A}},e^{-2t_{B}})P^{T}(\varphi)$.
Here, the matrix $P(\varphi)$ is defined in Eq.~(\ref{UV}),
$\varphi\in[0,2\pi)$, and $t_{A,B}\in(-\infty,+\infty)$. Hence,
from Eq.~(\ref{mathcalQ}) it then follows that
$\mathcal{Q}=\tilde{\alpha}$. Up to now we have considered only
CMs $\Gamma_{E}'$ built of the invertible matrices
$\tilde{\alpha}$. The remaining case of singular matrices
$\tilde{\alpha}$, which is also admissible, is recovered if we
take in the latter equality the limit $t_{A}\rightarrow+\infty$
and/or $t_{B}\rightarrow+\infty$. Consequently, the matrix
$\mathcal{Q}$ can be an arbitrary positive semidefinite matrix and
therefore instead of minimizing the RHS of Eq.~(\ref{K4}) over all
two-mode CMs $\Gamma_{E}'$ we can just carry out the minimization
over all $2\times 2$ real symmetric positive semidefinite matrices
$\mathcal{Q}$.

As we have already mentioned, such a matrix can be expressed as
$\mathcal{Q}=P(\varphi)\mbox{diag}(\lambda_{1},\lambda_{2})P^{T}(\varphi)$,
where $\varphi\in[0,2\pi)$ and $\lambda_{1,2}\geq0$. By
multiplying the matrices in the latter definition one finds after
some algebra that
\begin{eqnarray}\label{mathcalQ2}
\mathcal{Q}(\varphi,\lambda_{1},\lambda_{2})&=&\left(\begin{array}{cc}
L_{+}+L_{-}\cos(2\varphi) & L_{-}\sin(2\varphi) \\
L_{-}\sin(2\varphi) &  L_{+}-L_{-}\cos(2\varphi) \\
\end{array}\right),\nonumber\\
\end{eqnarray}
where $L_{\pm}=(\lambda_{1}\pm \lambda_{2})/2$ and where we have
written down explicitly the dependence of the matrix $\mathcal{Q}$
on the variables $\varphi, \lambda_{1}$ and $\lambda_{2}$. From equality
$\mathcal{Q}(\varphi+\pi,\lambda_{1},\lambda_{2})=\mathcal{Q}(\varphi,\lambda_{1},\lambda_{2})$
it further follows that we can assume without loss of generality
$\varphi\in[0,\pi)$ whereas equality
$\mathcal{Q}(\varphi+\frac{\pi}{2},\lambda_{1},\lambda_{2})=\mathcal{Q}(\varphi,\lambda_{2},\lambda_{1})$
allows us to assume $\lambda_{1}\geq\lambda_{2}$ in what follows.

Before performing minimization of the RHS of Eq.~(\ref{K4}) over
all matrices (\ref{mathcalQ2}), we first express the inverse
matrices on the RHS as
\begin{eqnarray}\label{mathcalQinv}
\left(\frac{1}{\nu}\openone+\mathcal{Q}\right)^{-1}&=&\frac{\frac{1}{\nu}\openone+J\mathcal{Q}J^{T}}{\tilde{d}},\nonumber\\
(\nu\openone+\mathcal{Q})^{-1}&=&\frac{\nu\openone+J\mathcal{Q}J^{T}}{d}
\end{eqnarray}
where matrix $J$ is defined in Eq.~(\ref{Omega}) and
\begin{eqnarray}\label{dtilded}
\tilde{d}&\equiv&\left(\frac{1}{\nu}+\lambda_{1}\right)\left(\frac{1}{\nu}+\lambda_{2}\right),\nonumber\\
d&\equiv&(\nu+\lambda_{1})(\nu+\lambda_{2}).
\end{eqnarray}
Substituting further into the obtained formula for matrix
$\mathcal{Q}$ from Eq.~(\ref{mathcalQ2}) and using
Eqs.~(\ref{eAB}) and (\ref{dtilded}) as well as identities
\begin{eqnarray}\label{zequalities}
\frac{z^{4}-1}{z^{4}+1}=\frac{k}{a},\quad
\frac{2z^{2}}{z^{4}+1}=\frac{\nu}{a},
\end{eqnarray}
one gets
\begin{eqnarray}\label{K5}
\mathscr{K}_{\rm h}&=&\frac{\left\{E+F\left[k\cos(2\varphi)-\nu\sin(2\varphi)\right]\right\}}{h(\lambda_{1})}\nonumber\\
&&\times\frac{\left\{E+F\left[k\cos(2\varphi)+\nu\sin(2\varphi)\right]\right\}}{h(\lambda_{2})}.\nonumber\\
\end{eqnarray}
Here,
\begin{eqnarray}
E&\equiv&1+\lambda_{1}\lambda_{2}+\cosh(v)(\lambda_{1}+\lambda_{2}),\label{E}\\
F&\equiv&\frac{\sinh(v)}{a}(\lambda_{1}-\lambda_{2}),\label{F}
\end{eqnarray}
where we have set $\cosh(v)=(\nu+1/\nu)/2$, and
\begin{equation}\label{hj}
h(\lambda_{j})\equiv1+2\cosh(v)\lambda_{j}+\lambda_{j}^{2},\quad
j=1,2.
\end{equation}
Finally, by multiplying the expressions in the curly brackets on
the RHS of equation (\ref{K5}) we get the quantity to be minimized
in the following very simple form
\begin{eqnarray}\label{K6}
\mathscr{K}_{\rm
h}=\frac{a^2-k^2}{a^2}+\mathcal{K}(\varphi,\lambda_{1},\lambda_{2})
\end{eqnarray}
with
\begin{eqnarray}\label{mathcalK}
\mathcal{K}(\varphi,\lambda_{1},\lambda_{2})\equiv\frac{\left[\frac{k}{a}E+\tilde{F}\cos(2\varphi)\right]^2}{E^{2}-\tilde{F}^2},
\end{eqnarray}
where we have introduced
\begin{equation}\label{tildeF}
\tilde{F}\equiv Fa=\sinh(v)(\lambda_{1}-\lambda_{2}),
\end{equation}
and where we have used identities $E^{2}-\tilde{F}^2=h(\lambda_{1})h(\lambda_{2})$ and (\ref{nuis}).

Having simplified function $\mathscr{K}_{\rm h}$ we can now proceed
with the minimization on the RHS of Eq.~(\ref{Kmin}). We have already shown
that the minimization over all CMs $\Gamma_{E}$ can be replaced with the
minimization with respect to all positive semidefinite matrices (\ref{mathcalQ2}).
From Eq.~(\ref{K6}) it then follows that
\begin{equation}\label{Kmin2}
\mathscr{K}_{\rm min}=\frac{a^2-k^2}{a^2}+\mathcal{K}_{\rm min}
\end{equation}
with
\begin{equation}\label{mathcalKmin}
\mathcal{K}_{\rm
min}\equiv\inf_{\substack{\varphi\in[0,\pi)\\
    \lambda_{1}\geq\lambda_{2}\geq0}}\mathcal{K}(\varphi,\lambda_{1},\lambda_{2}),
\end{equation}
where the function $\mathcal{K}$ is defined in
Eq.~(\ref{mathcalK}). Our task is thus to minimize function
(\ref{mathcalK}) on a subset ($\equiv\mathcal{O}$) of the
three-dimensional Euclidean space of the variables $\lambda_{1},
\lambda_{2}$ and $\varphi$, which is characterized by inequalities
$\lambda_{2}\geq0, \lambda_{1}\geq\lambda_{2}$ and
$0\leq\varphi<\pi$.

We carry out the minimization in Eq.~(\ref{mathcalKmin}) by
finding a lower bound on function (\ref{mathcalK}) which is
independent of the variables we minimize over, and which is tight,
i.e., the bound is reached at some point from the set
$\mathcal{O}$. In order to find the bound note first, that for
$\lambda_{1}>\lambda_{2}$ one gets $\tilde{F}>0$. From inequality
$\cos(2\varphi)\geq -1$ it then follows that
$\tilde{F}\cos(2\varphi)\geq -\tilde{F}$ and hence
\begin{equation}\label{EFineq1}
\frac{k}{a}E+\tilde{F}\cos(2\varphi)\geq\frac{k}{a}E-\tilde{F}.
\end{equation}
Using Eqs.~(\ref{E}) and (\ref{tildeF}) one further finds that
\begin{eqnarray}\label{fracEtildeF}
\frac{E}{\tilde{F}}&=&2\frac{(1+\lambda_{1}\lambda_{2})\sqrt{a^2-k^2}+\lambda_{2}(a^2-k^2+1)}{(\lambda_{1}-\lambda_{2})(a^2-k^2-1)}\nonumber\\
&&+1+\frac{2}{a^2-k^2-1}\geq1+\frac{2}{a^2-k^2-1}\nonumber\\
&>&\frac{a}{a-1}>\frac{a}{k},
\end{eqnarray}
where the second and the third inequality follows from the
necessary and sufficient condition for entanglement of a symmetric
squeezed thermal state, $1>a-k$. The latter inequality implies
fulfilment of the following inequality $(k/a)E-\tilde{F}>0$, which
allows us to show easily using inequality (\ref{EFineq1}) that the
function (\ref{mathcalK}) is lower bounded as
\begin{eqnarray}\label{Klowerbound}
\mathcal{K}=\frac{\left[\frac{k}{a}E+\tilde{F}\cos(2\varphi)\right]^2}{E^{2}-\tilde{F}^2}\geq\frac{\left(\frac{k}{a}E-\tilde{F}\right)^2}{E^{2}-\tilde{F}^2}\equiv\mathcal{L},\nonumber\\
\end{eqnarray}
where we did not write explicitly the dependence of the function
$\mathcal{K}$ on variables $\lambda_{1}, \lambda_{2}$ and
$\varphi$ for brevity.

To proceed further, it is now convenient to express the lower
bound $\mathcal{L}$ in terms of the ratio
$\mathcal{R}\equiv\tilde{F}/E$ as
\begin{eqnarray}\label{L}
\mathcal{L}=\frac{\left(\frac{k}{a}-\mathcal{R}\right)^2}{1-\mathcal{R}^2}.
\end{eqnarray}
Calculating now the derivative of the latter function with respect
to the variable $\mathcal{R}$ we find that it holds
\begin{eqnarray}\label{derL}
\frac{d\mathcal{L}}{d\mathcal{R}}=-2\frac{\left(\frac{k}{a}-\mathcal{R}\right)\left(1-\frac{k}{a}\mathcal{R}\right)}{\left(1-\mathcal{R}^2\right)^{2}}<0,
\end{eqnarray}
as follows from inequalities $k/a-\mathcal{R}>0$ and
$1-(k/a)\mathcal{R}>0$. Here, the last inequality can be derived
straightforwardly using inequality (\ref{fracEtildeF}),
$E/\tilde{F}>a/k$, and the state condition $a^2-k^2\geq1$.
Inequality (\ref{derL}) reveals, that function $\mathcal{L}$ is a
strictly decreasing function of the ratio $\mathcal{R}$ and
therefore it is minimized if $\mathcal{R}$ attains the largest
value ($\equiv\mathcal{R}_{\max}$) on the set $\mathcal{O}$. The
maximum value $\mathcal{R}_{\max}$ can be found by minimization of
the inverse ratio $1/\mathcal{R}=E/\tilde{F}$ which can be done
with the help of inequalities (\ref{fracEtildeF}). Namely, the
first of the inequalities unveils that
\begin{eqnarray}\label{Rmax}
\mathcal{R}_{\max}=\frac{1}{1+\frac{2}{a^2-k^2-1}}
\end{eqnarray}
and it is attained for $\lambda_{1}=e^{2t}$ and
$\lambda_{2}=e^{-2t}$ in the limit for $t\rightarrow+\infty$. This
then gives
\begin{eqnarray}\label{Kmin3}
\mathcal{K}_{\rm min}=\frac{\left(\frac{k}{a}-\mathcal{R}_{\rm
max}\right)^2}{1-\mathcal{R}_{\rm max}^2}=\frac{a^{2}-k^{2}}{a^2}\left(\frac{a-k-\frac{1}{a-k}}{2}\right)^2\nonumber\\
\end{eqnarray}
and hence for function (\ref{Kmin2}) one gets
\begin{equation}\label{Kminfinal}
\mathscr{K}_{\rm
min}=\frac{a^{2}-k^{2}}{a^2}\left[\frac{(a-k)^2+1}{2(a-k)}\right]^2.
\end{equation}
Inserting finally from here into Eq.~(\ref{Uboundis2}) we get the
following analytical formula for the upper bound (\ref{Uboundis}):
\begin{equation}\label{Uboundisfinal}
U\left(\rho_{AB}^{(2)}\right)=\ln\left[\frac{(a-k)^2+1}{2(a-k)}\right].
\end{equation}

It remains to investigate the behavior of the function
(\ref{mathcalK}) for $\lambda_{1}=\lambda_{2}$. Here, we get
$\tilde{F}=0$ and hence
$\mathcal{K}_{\lambda_{1}=\lambda_{2}}=k^2/a^2$ according to
Eq.~(\ref{mathcalK}). From Eq.~(\ref{Kmin2}) it then follows that
$\mathscr{K}_{\lambda_{1}=\lambda_{2}}\equiv(a^2-k^2)/a^2+\mathcal{K}_{\lambda_{1}=\lambda_{2}}=1$
and hence $U_{\lambda_{1}=\lambda_{2}}(\rho_{AB}^{(2)})\equiv
\frac{1}{2}\ln[a^2/(a^2-k^2)]+\frac{1}{2}\ln\mathscr{K}_{\lambda_{1}=\lambda_{2}}=\ln(a/\sqrt{a^2-k^2})$
holds by Eq.~(\ref{Uboundis2}). Let us now consider the following
expressions $U(\rho_{AB}^{(2)})=\ln[\cosh(p_1)]$ and
$U_{\lambda_{1}=\lambda_{2}}(\rho_{AB}^{(2)})=\ln[\cosh(p_2)]$,
where the parameters $p_{1}$ and $p_{2}$ are defined via equations
$e^{p_{1}}=1/(a-k)$ and $e^{p_{2}}=\sqrt{(a+k)/(a-k)}$. Then we
see from state condition $a^2-k^2\geq1$ that $e^{p_{2}}\geq
e^{p_{1}}$ and therefore $p_{2}\geq p_{1}$. Consequently,
$U_{\lambda_{1}=\lambda_{2}}(\rho_{AB}^{(2)})\geq
U(\rho_{AB}^{(2)})$ and Eq.~(\ref{Uboundisfinal}) is really the
sought least upper bound defined in Eq.~(\ref{Uboundis}).

Like in the case of symmetric GLEMS the formula
(\ref{Uboundisfinal}) represents an upper bound on the GIE for a
two-mode symmetric squeezed thermal state for which the parameter
$a$ satisfies inequality (\ref{Finalupperbound2}). Because the
upper bound represents for homodyne detections of quadratures
$x_{A}$ and $x_{B}$ on modes $A$ and $B$ of the minimal
purification of the state $\rho_{AB}^{(2)}$ the least classical
mutual information over all Gaussian measurements on the purifying
subsystem $E$ it is the largest minimal mutual information which
can be achieved. Thus the upper bound (\ref{Uboundisfinal})
coincides with the sought GIE and therefore GIE for two-mode
symmetric squeezed thermal states satisfying inequality
(\ref{Finalupperbound2}) is given by
\begin{equation}\label{GIEisfinal}
E_{\downarrow}^{G}\left(\rho_{AB}^{(2)}\right)=\ln\left[\frac{(a-k)^2+1}{2(a-k)}\right].
\end{equation}

From our previous analysis it follows, that the matrix
$\mathcal{Q}$ for which the GIE is achieved, is obtained from the
matrix
$\mathcal{Q}(\pi/2,\lambda_{1}=e^{2t},\lambda_{2}=e^{-2t})=\mbox{diag}(e^{-2t},e^{2t})$
in the limit for $t\rightarrow+\infty$. This reveals, that it is
optimal for Eve to carry out homodyne detection of quadrature
$x_{E_{A}}$ and $p_{E_{B}}$ on her modes $E_{A}$ and $E_{B}$,
respectively, which is described by CM
$\Gamma_{E_{A}E_{B}}^{t}=\mbox{diag}(e^{-2t},e^{2t})_{E_{A}}\oplus\mbox{diag}(e^{2t},e^{-2t})_{E_{B}}$
in the limit for $t\rightarrow+\infty$.

\section{GIE for asymmetric states}\label{Sec_II}

Up to now, GIE was calculated for several subclasses of symmetric
states described by CM (\ref{gammast2}) with $a=b$. In
Ref.~\cite{Mista_15} GIE was calculated for all pure states,
whereas Sec.~\ref{Sec_I} contains derivation of GIE for all
symmetric GLEMS as well as for all symmetric squeezed thermal
states satisfying condition $a\leq2.41$.

In this section, we extend the analysis of GIE to states with CM
(\ref{gammast2}), where parameters $a$ and $b$ may differ.
Specifically, we focus on generally asymmetric squeezed thermal
states which satisfy condition $k_{x}=k_{p}\equiv k$, i.e., their
CM reads as
\begin{eqnarray}\label{gammast3}
\gamma_{AB}=\left(\begin{array}{cccc}
a & 0 & k & 0\\
0 & a & 0 & -k \\
k & 0 & b & 0 \\
0 & -k & 0 & b \\
\end{array}\right).
\end{eqnarray}
Insertion of the latter condition into the RHS of Eq.~(\ref{nu12})
yields the symplectic eigenvalues in the form
\begin{equation}\label{nu12tmsts}
\nu_{1,2}=\frac{\sqrt{(a+b)^{2}-4k^2}\pm|a-b|}{2}.
\end{equation}
Also the symplectic matrix $S$, which brings CM (\ref{gammast3})
to the Williamson normal form (\ref{Williamson}), is quite simple
as it is given by a matrix of a two-mode squeezer. Indeed, if
$a\geq b$, then it is not difficult to show, that symplectic
matrix
\begin{eqnarray}\label{Stmsts}
S=\left(\begin{array}{cc}
x\openone & -y\sigma_{z}\\
-y\sigma_{z} & x\openone\\
\end{array}\right)
\end{eqnarray}
with
\begin{eqnarray}\label{xytmsts}
x&=&\sqrt{\frac{a+b+\sqrt{(a+b)^2-4k^2}}{2\sqrt{(a+b)^{2}-4k^2}}},\nonumber\\
y&=&\sqrt{\frac{a+b-\sqrt{(a+b)^2-4k^2}}{2\sqrt{(a+b)^{2}-4k^2}}}
\end{eqnarray}
brings CM (\ref{gammast3}) to the Williamson normal form, where
the symplectic eigenvalues are given in Eq.~(\ref{nu12tmsts}). If,
on the other hand, $a<b$, then symplectic matrix
\begin{eqnarray}\label{tildeStmsts}
\tilde{S}=TS=\left(\begin{array}{cc}
-y\sigma_{z} & x\openone\\
x\openone & -y\sigma_{z}\\
\end{array}\right),
\end{eqnarray}
where
\begin{equation}\label{T}
T=\left(\begin{array}{cc}
0 & \openone \\
\openone & 0\\
\end{array}\right),
\end{equation}
brings CM (\ref{gammast3}) to the Williamson normal form
(\ref{Williamson}), where the symplectic eigenvalues are again
given in Eq.~(\ref{nu12tmsts}). Further, since parameters
(\ref{xytmsts}) satisfy equation $x^2-y^2=1$, one can introduce
new parametrization $x=\cosh(r)$ and $y=\sinh(r)$, $r>0$, which
reveals that $S$ is really nothing but a symplectic matrix of a
two-mode squeezer with a squeezing parameter $r$. As for
symplectic matrix $\tilde{S}$, it is just a symplectic matrix of
the same squeezer followed by a completely reflecting beam
splitter exchanging modes $A$ and $B$, which is described by
symplectic matrix (\ref{T}).

\subsection{GIE for asymmetric GLEMS}\label{Subsec_asymGLEMS}

This subsection contains calculation of GIE for states with CM
(\ref{gammast3}) which are GLEMS, i.e., their lower symplectic
eigenvalue is equal to $\nu_{2}=1$, which we shall further denote
as $\rho_{AB}^{(3)}$. From Eqs.~(\ref{nu12tmsts}) and $\nu_{2}=1$
one finds easily that the parameter $k$ is then given by
\begin{equation}\label{k}
k=\left\{\begin{array}{lll} \sqrt{(a+1)(b-1)}, & \textrm{if} & a\geq b;\\
\sqrt{(a-1)(b+1)}, & \textrm{if}  & a<b,
\end{array}\right.
\end{equation}
and consequently CM (\ref{gammast3}) attains for $a\geq b$ the
form
\begin{eqnarray}\label{GLEMS}
\bar{\gamma}_{AB}^{(3)}=\left(\begin{array}{cc}
a\openone & \sqrt{(a+1)(b-1)}\sigma_{z} \\
\sqrt{(a+1)(b-1)}\sigma_{z} & b\openone \\
\end{array}\right),\nonumber\\
\end{eqnarray}
and for $a<b$ the form
\begin{eqnarray}\label{tildeGLEMS}
\tilde{\gamma}_{AB}^{(3)}=\left(\begin{array}{cc}
a\openone & \sqrt{(a-1)(b+1)}\sigma_{z} \\
\sqrt{(a-1)(b+1)}\sigma_{z} & b\openone \\
\end{array}\right).\nonumber\\
\end{eqnarray}
Apart from lower symplectic eigenvalue $\nu_{2}=1$ both CMs
(\ref{GLEMS}) and (\ref{tildeGLEMS}) possesses the larger
symplectic eigenvalue equal to
\begin{equation}\label{nu1GLEMS}
\tilde{\nu}\equiv\nu_{1}=1+|a-b|.
\end{equation}
Hence we see, that if $a=b$, we get $\nu_{1}=1$ and sice also
$\nu_{2}=1$ CMs (\ref{GLEMS}) and (\ref{tildeGLEMS}) describe a
pure state. As the case of pure states was already analyzed in
Ref.~\cite{Mista_15}, we will restrict ourself only to CMs for
which parameters $a$ and $b$ satisfy strict inequalities $a>b$ and
$a<b$.

Let us first analyze the case when $a>b$. Since $\nu_{1}>1$
whereas $\nu_{2}=1$ we have $R=1$. For CM (\ref{gammacondS}) we
then get expression as in Eq.~(\ref{gammacondS2}),
\begin{equation}\label{gammacondGLEMS3}
\bar{\gamma}_{AB|E}^{(3)}=S^{-1}\left[\gamma_{A|E}(\tilde{\nu})\oplus\openone_{B}\right]\left(S^{-1}\right)^{T},
\end{equation}
in which
\begin{equation}\label{gammaAEtilde}
\gamma_{A|E}(\tilde{\nu})=P^{T}(\varphi)\mbox{diag}(\tilde{\mathcal{V}}_{x},\tilde{\mathcal{V}}_{p})P(\varphi),
\end{equation}
where
\begin{equation}\label{tildecalVxVp}
\tilde{\mathcal{V}}_{x}=\frac{\tilde{\nu}
V_{x}+1}{\tilde{\nu}+V_{x}},\quad
\tilde{\mathcal{V}}_{p}=\frac{\tilde{\nu}
V_{p}+1}{\tilde{\nu}+V_{p}}
\end{equation}
and $S$ is the symplectic matrix (\ref{Stmsts}) with
\begin{equation}\label{xy}
x=\sqrt{\frac{a+1}{a-b+2}},\quad y=\sqrt{\frac{b-1}{a-b+2}}.
\end{equation}
Making use of expressions $x=\cosh(r)$ and $y=\sinh(r)$, and
replacing $r$ with $-r$ or using formula $S^{-1}=\Omega
S^{T}\Omega^{T}$, we get an inverse symplectic matrix $S^{-1}$
appearing on the RHS of Eq.~(\ref{gammacondGLEMS3}). Insertion of
the latter matrix and CM $\gamma_{A|E}(\tilde{\nu})$ into the RHS
of Eq.~(\ref{gammacondGLEMS3}) yields the sought CM
\begin{equation}\label{gammacondSGLEMS}
\bar{\gamma}_{AB|E}^{(3)}=[P^{T}(\varphi)\oplus
P(\varphi)]\delta_{AB|E}[P(\varphi)\oplus P^{T}(\varphi)],
\end{equation}
where
\begin{eqnarray}\label{deltacond}
\delta_{AB|E}=\left(\begin{array}{cccc}
\alpha_{x} & 0 & \tau_{x} & 0\\
0 & \alpha_{p} & 0 & -\tau_{p} \\
\tau_{x} & 0 & \beta_{x} & 0 \\
0 & -\tau_{p} & 0 & \beta_{p} \\
\end{array}\right),
\end{eqnarray}
where
\begin{eqnarray}\label{alphabetatau}
\alpha_{x,p}&=&x^2\tilde{\mathcal{V}}_{x,p}+y^2,\nonumber\\
\beta_{x,p}&=&y^2\tilde{\mathcal{V}}_{x,p}+x^2,\nonumber\\
\tau_{x,p}&=&xy(\tilde{\mathcal{V}}_{x,p}+1).
\end{eqnarray}
For further use we need finally to bring CM
(\ref{gammacondSGLEMS}) to the standard form. Since the CM and CM
(\ref{deltacond}) are connected by local symplectic
transformation, they possess the same standard form and therefore
it suffices to bring CM (\ref{deltacond}) to the standard form.
Application of local squeezing transformations
$\mbox{diag}(\lambda_{A}^{-1},\lambda_{A})\oplus\mbox{diag}(\lambda_{B}^{-1},\lambda_{B})$,
where
\begin{eqnarray}\label{lambdaAB}
\lambda_{A}=\sqrt[4]{\frac{x^2\tilde{\mathcal{V}}_{x}+y^2}{x^2\tilde{\mathcal{V}}_{p}+y^2}},\quad
\lambda_{B}=\sqrt[4]{\frac{y^2\tilde{\mathcal{V}}_{x}+x^2}{y^2\tilde{\mathcal{V}}_{p}+x^2}},
\end{eqnarray}
does the job and brings CM (\ref{deltacond}) to standard form
(\ref{gammast2}), where parameters $a,b,k_{x}$ and $k_{p}$ are
replaced with
\begin{eqnarray}\label{tildeabkxkp}
\tilde{a}&=&\sqrt{(x^2\tilde{\mathcal{V}}_{x}+y^2)(x^2\tilde{\mathcal{V}}_{p}+y^2)},\nonumber\\
\tilde{b}&=&\sqrt{(y^2\tilde{\mathcal{V}}_{x}+x^2)(y^2\tilde{\mathcal{V}}_{p}+x^2)},\nonumber\\
\tilde{k}_{x}&=&\frac{xy(\tilde{\mathcal{V}}_{x}+1)}{\lambda_{A}\lambda_{B}},\quad
\tilde{k}_{p}=xy(\tilde{\mathcal{V}}_{p}+1)\lambda_{A}\lambda_{B}.
\nonumber\\
\end{eqnarray}
Consequently, parameters (\ref{tildeabkxkp}) are also parameters
of the standard form of CM $\bar{\gamma}_{AB|E}^{(3)}$ as needed.

To evaluate GIE for states ($\equiv\bar{\rho}_{AB}^{(3)}$) with CM
(\ref{GLEMS}), we now use the method of Sec.~\ref{Sec_I}. In this
method, instead of carrying out directly optimization in the
definition of GIE, Eq.~(\ref{GIE}), we first calculate an easier
computable upper bound, Eq.~(\ref{Ubound}),
\begin{equation}\label{Uboundhelp}
U\left(\bar{\rho}_{AB}^{(3)}\right)=\mathop{\mbox{inf}}_{\Gamma_{E}}\left[\mathcal{I}_{c}^{G}\left(\bar{\rho}_{AB|E}^{(3)}\right)\right],
\end{equation}
where $\mathcal{I}_{c}^{G}(\bar{\rho}_{AB|E}^{(3)})$ is GCMI
(\ref{IcG}) of conditional state $\bar{\rho}_{AB|E}^{(3)}$ with CM
(\ref{gammacondSGLEMS}). For CM $\bar{\gamma}_{AB|E}^{(3)}$ for
which inequality (\ref{IcGgencond}) holds, where parameter
$\tilde{c}_{x}$ is replaced with parameter $\tilde{k}_{x}$, i.e.,
\begin{equation}\label{IcGcond}
\sqrt{\frac{\tilde{a}}{\tilde{b}}}+\sqrt{\frac{\tilde{b}}{\tilde{a}}}+\frac{1}{\sqrt{\tilde{a}\tilde{b}}}-\sqrt{\tilde{a}\tilde{b}-\tilde{k}_{x}^2}\geq0,
\end{equation}
the GCMI is given by
\begin{equation}\label{IcG1}
\mathcal{I}_{c}^{G}\left(\bar{\rho}_{AB|E}^{(3)}\right)=\frac{1}{2}\ln\frac{1}{1-h},
\end{equation}
where
\begin{equation}\label{h}
h=\frac{\tilde{k}_{x}^2}{\tilde{a}\tilde{b}},
\end{equation}
and it is reached by homodyne detection of quadratures $x_{A}$ and
$x_{B}$ on modes $A$ and $B$. If we now substitute to the RHS from
Eqs.~(\ref{tildeabkxkp}) one finds, that
\begin{equation}\label{h1}
h=\frac{x^{2}y^{2}(\tilde{\mathcal{V}}_{x}+1)^{2}}{(x^2\tilde{\mathcal{V}}_{x}+y^2)(y^2\tilde{\mathcal{V}}_{x}+x^2)}=\frac{1}{1+\frac{\tilde{\mathcal{V}}_{x}}{x^{2}y^{2}(\tilde{\mathcal{V}}_{x}+1)^{2}}},
\end{equation}
where we have used equality $x^2-y^2=1$. Provided that inequality
(\ref{IcGcond}) is satisfied for any CM $\Gamma_{E}$, the upper
bound (\ref{Ubound}) can be calculated by minimization of GCMI
(\ref{IcG1}) over all CMs $\Gamma_{E}$. This amounts to
minimization of quantity (\ref{h1}) over all admissible values of
the parameter $\tilde{\mathcal{V}}_{x}$ defined in
Eq.~(\ref{tildecalVxVp}), which is equivalent with maximization of
the function
\begin{equation}\label{calF}
\mathcal{F}(\tilde{\mathcal{V}}_{x})\equiv\frac{\tilde{\mathcal{V}}_{x}}{(\tilde{\mathcal{V}}_{x}+1)^{2}}.
\end{equation}
Equation (\ref{tildecalVxVp}) reveals, that the parameter
$\tilde{\mathcal{V}}_{x}$ lies in the interval
$\tilde{\mathcal{V}}_{x}\in[1,\tilde{\nu}]$ on which function
(\ref{calF}) is monotonically decreasing. Thus $\mathcal{F}$ is
maximized by $\tilde{\mathcal{V}}_{x}=1$, which corresponds to
heterodyne detection on Eve's mode $E$, i.e., a projection of the
mode onto a coherent state. Then we get
$\mathcal{F}(\tilde{\mathcal{V}}_{x}=1)=1/4$ which further gives
\begin{equation}\label{hmin}
h_{\rm min}\equiv
h(\tilde{\mathcal{V}}_{x}=1)=1-\frac{1}{1+4x^{2}y^{2}}=1-\frac{(a-b+2)^{2}}{(a+b)^{2}},
\end{equation}
where in the derivation of the second equality we have used
equality $1+4x^{2}y^{2}=(a+b)^{2}/(a-b+2)^{2}$. Insertion of the
minimal value of quantity (\ref{h}) given in Eq.~(\ref{hmin}) into
the RHS of Eq.~(\ref{IcG1}) yields finally the upper bound
\begin{equation}\label{UboundGLEMS}
U\left(\bar{\rho}_{AB}^{(3)}\right)=\ln\left(\frac{a+b}{a-b+2}\right).
\end{equation}
Viewed from a different perspective, the RHS of the latter
equality represents for homodyne detection of quadratures $x_{A}$
and $x_{B}$ on modes $A$ and $B$ the least mutual information
(\ref{f}) over all Gaussian measurements on mode $E$. Since the
mutual information saturates the upper bound, no Gaussian
measurements on modes $A$ and $B$ can give a larger mutual
information minimized over all CMs $\Gamma_{E}$, and GIE thus
coincides with the upper bound (\ref{UboundGLEMS}). Therefore, for
asymmetric GLEMS with CM (\ref{GLEMS}) for which inequality
(\ref{IcGcond}) is satisfied for any CM $\Gamma_{E}$, GIE reads as
\begin{equation}\label{GIEGLEMS}
E_{\downarrow}^{G}\left(\bar{\rho}_{AB}^{(3)}\right)=\ln\left(\frac{a+b}{a-b+2}\right).
\end{equation}

It remains to identify the set of states with CM (\ref{GLEMS}),
where $a>b$, for which inequality (\ref{IcGcond}) is satisfied for
any CM $\Gamma_{E}$ and therefore GIE is given by formula
(\ref{GIEGLEMS}). Unfortunately, inequality (\ref{IcGcond}) is
very complex which makes identification of the whole set a
difficult task. Here, instead of discussing inequality
(\ref{IcGcond}), we therefore use a much more simple inequality
(\ref{sqrtabbound}) the fulfilment of which suffices for original
inequality (\ref{IcGcond}) to hold. If we now substitute for
$\tilde{a}$ and $\tilde{b}$ on the LHS of the inequality
(\ref{sqrtabbound}) from Eq.~(\ref{tildeabkxkp}), we get the
following relations
\begin{eqnarray}\label{tildeabbound1}
\sqrt{\tilde{a}\tilde{b}}&=&\sqrt[4]{(x^2\tilde{\mathcal{V}}_{x}+y^2)(x^2\tilde{\mathcal{V}}_{p}+y^2)}\nonumber\\
&&\times\sqrt[4]{(y^2\tilde{\mathcal{V}}_{x}+x^2)(y^2\tilde{\mathcal{V}}_{p}+x^2)}\nonumber\\
&\leq&\sqrt{(x^2\tilde{\nu}+y^2)(y^2\tilde{\nu}+x^2)}\nonumber\\
&=&\sqrt{ab},
\end{eqnarray}
where the first inequality follows from inequalities
$\tilde{\mathcal{V}}_{x,p}\leq\tilde{\nu}$ and for derivation of
the second equality we used formulas (\ref{nu1GLEMS}) and
(\ref{xy}). Hence we see, that if
\begin{equation}\label{abbound}
\sqrt{ab}\leq2.41,
\end{equation}
then also inequality (\ref{sqrtabbound}) is fulfilled. Thus we
have arrived to the finding that for all GLEMS with CM
(\ref{GLEMS}), where $a>b$, which satisfy inequality
(\ref{abbound}) GIE is given by formula (\ref{GIEGLEMS}).

Moving to the case of states ($\equiv\tilde{\rho}_{AB}^{(3)}$)
with CM (\ref{tildeGLEMS}), where $a<b$, Eqs.~(\ref{xytmsts}) and
(\ref{tildeStmsts}) reveal that the symplectic matrix which brings
CM (\ref{tildeGLEMS}) to the Williamson normal form
(\ref{Williamson}) reads as
\begin{eqnarray}\label{tildeSymplGLEMS}
\tilde{S}=\left(\begin{array}{cc}
-\tilde{y}\sigma_{z} & \tilde{x}\openone\\
\tilde{x}\openone & -\tilde{y}\sigma_{z}\\
\end{array}\right),
\end{eqnarray}
where
\begin{equation}\label{tildexy}
\tilde{x}=\sqrt{\frac{b+1}{b-a+2}},\quad
\tilde{y}=\sqrt{\frac{a-1}{b-a+2}}.
\end{equation}
The CM (\ref{gammacondS}) ($\equiv\tilde{\gamma}_{AB|E}^{(3)}$) is
then obtained by replacing in Eq.~(\ref{gammacondGLEMS3}) matrix
$S$ with matrix $\tilde{S}$, Eq.~(\ref{tildeSymplGLEMS}). Further,
the CM can be brought to the form
\begin{equation}\label{gammacondSGLEMS2}
\tilde{\gamma}_{AB|E}^{(3)}=[P(\varphi)\oplus
P^{T}(\varphi)]\tilde{\delta}_{AB|E}[P^{T}(\varphi)\oplus
P(\varphi)]
\end{equation}
with
\begin{eqnarray}\label{tildedeltacond}
\tilde{\delta}_{AB|E}=\left(\begin{array}{cccc}
\tilde{\beta}_{x} & 0 & \tilde{\tau}_{x} & 0\\
0 & \tilde{\beta}_{p} & 0 & -\tilde{\tau}_{p} \\
\tilde{\tau}_{x} & 0 & \tilde{\alpha}_{x} & 0 \\
0 & -\tilde{\tau}_{p} & 0 & \tilde{\alpha}_{p} \\
\end{array}\right),
\end{eqnarray}
where parameters $\tilde{\alpha}_{x,p}, \tilde{\beta}_{x,p}$ and
$\tilde{\tau}_{x,p}$ are given by formulas (\ref{alphabetatau}),
in which parameters $x$ and $y$ on the RHS are replaced with
parameters $\tilde{x}$ and $\tilde{y}$ defined in
Eq.~(\ref{tildexy}). By local squeezing transformations
$\mbox{diag}(\tilde{\lambda}_{B}^{-1},\tilde{\lambda}_{B})\oplus\mbox{diag}(\tilde{\lambda}_{A}^{-1},\tilde{\lambda}_{A})$,
where
\begin{eqnarray}\label{tildelambdaAB}
\tilde{\lambda}_{A}=\sqrt[4]{\frac{\tilde{x}^2\tilde{\mathcal{V}}_{x}+\tilde{y}^2}{\tilde{x}^2\tilde{\mathcal{V}}_{p}+\tilde{y}^2}},\quad
\tilde{\lambda}_{B}=\sqrt[4]{\frac{\tilde{y}^2\tilde{\mathcal{V}}_{x}+\tilde{x}^2}{\tilde{y}^2\tilde{\mathcal{V}}_{p}+\tilde{x}^2}},
\end{eqnarray}
one can further transform CM (\ref{tildedeltacond}) to standard
form (\ref{gammast2}), where parameters $a,b,k_{x}$ and $k_{p}$
are replaced with parameters
\begin{eqnarray}\label{primedabkxkp}
a'&=&\sqrt{(\tilde{y}^2\tilde{\mathcal{V}}_{x}+\tilde{x}^2)(\tilde{y}^2\tilde{\mathcal{V}}_{p}+\tilde{x}^2)},\nonumber\\
b'&=&\sqrt{(\tilde{x}^2\tilde{\mathcal{V}}_{x}+\tilde{y}^2)(\tilde{x}^2\tilde{\mathcal{V}}_{p}+\tilde{y}^2)},\nonumber\\
k_{x}'&=&\frac{\tilde{x}\tilde{y}(\tilde{\mathcal{V}}_{x}+1)}{\tilde{\lambda}_{A}\tilde{\lambda}_{B}},\quad
k_{p}'=\tilde{x}\tilde{y}(\tilde{\mathcal{V}}_{p}+1)\tilde{\lambda}_{A}\tilde{\lambda}_{B}.
\nonumber\\
\end{eqnarray}

Now it is easy to derive formula for GIE for states described by
CM (\ref{tildeGLEMS}), where $a<b$. Namely, the derivation is
exactly the same as the one which we have performed between
Eqs.(\ref{Uboundhelp}) and (\ref{abbound}). The only difference is
that parameters $\tilde{a}, \tilde{b}$ and $\tilde{k}_{x}$ are
replaced with parameters $a', b'$ and $k_{x}'$,
Eq.~(\ref{primedabkxkp}), and parameters $x$ and $y$ are replaced
with parameters $\tilde{x}$ and $\tilde{y}$, Eq.~(\ref{tildexy}).
In this way we get GIE for states with CM (\ref{tildeGLEMS}),
where $a<b$, in the form (\ref{GIEGLEMS}), where parameters $a$
and $b$ are exchanged, i.e.,
\begin{equation}\label{tildeGIEGLEMS}
E_{\downarrow}^{G}\left(\tilde{\rho}_{AB}^{(3)}\right)=\ln\left(\frac{a+b}{b-a+2}\right).
\end{equation}

 Consequently, for asymmetric GLEMS with CM
(\ref{gammast3}), where parameter $k$ is given in Eq.(\ref{k}),
and where parameters $a$ and $b$ satisfy inequality
(\ref{abbound}), GIE is equal to
\begin{equation}\label{GIEGLEMSfinal}
E_{\downarrow}^{G}\left(\rho_{AB}^{(3)}\right)=\ln\left(\frac{a+b}{|a-b|+2}\right).
\end{equation}


\section{Equality to Gaussian R\'{e}nyi-2 entanglement}\label{Sec_III}

Investigation of GIE carried out in Ref.~\cite{Mista_15} and the
main text revealed a remarkable relation of it to another
quantifier of Gaussian entanglement called Gaussian R\'{e}nyi-2
(GR2) entanglement \cite{Adesso_12}. Concretely, it was shown,
that for all pure two-mode Gaussian states, all symmetric GLEMS as
well as all symmetric squeezed thermal states satisfying condition
$a\leq 2.41$, GIE {\it is equal} to GR2 entanglement. A next
logical step is then to analyze relation of GIE and GR2
entanglement for asymmetric states. Here, we will show, that
equality of the two quantities holds also for all states
investigated in previous Appendix~\ref{Sec_II}, i.e., for
all asymmetric two-mode squeezed thermal states with a three-mode
purification, which satisfy inequality $\sqrt{ab}\leq2.41$.

At the outset, we will evaluate GR2 entanglement ($\equiv
E_{2}^{G}$) for states $\bar{\rho}_{AB}^{(3)}$ with CM
(\ref{GLEMS}), where $a>b$. This can be done using an analytic
formula for GR2 entanglement of a two-mode reduction of a
three-mode pure state. The formula utilizes a standard form CM to
which any pure three-mode Gaussian state of modes $A_{1}, A_{2}$
and $A_{3}$ can be brought by local symplectic transformations
\begin{eqnarray}\label{threemodepure}
\gamma_{A_{1}A_{2}A_{3}}=\left(\begin{array}{cccccc}
a_{1} & 0 & c_{3}^{+} & 0 & c_{2}^{+} & 0\\
0 & a_{1} & 0 & c_{3}^{-} & 0 & c_{2}^{-}\\
c_{3}^{+} & 0 & a_{2} & 0 & c_{1}^{+} & 0\\
0 & c_{3}^{-} & 0 & a_{2} & 0 & c_{1}^{-}\\
c_{2}^{+} & 0 & c_{1}^{+} & 0 & a_{3} & 0\\
0 & c_{2}^{-} & 0 & c_{1}^{-} & 0 & a_{3}\\
\end{array}\right),
\end{eqnarray}
with
\begin{eqnarray}\label{ci}
c_{i}^{\pm}&=&\frac{\sqrt{a_{--}a_{+-}}\pm\sqrt{a_{-+}a_{++}}}{4\sqrt{a_{j}a_{k}}},
\end{eqnarray}
where
\begin{eqnarray}\label{apm}
a_{\mp-}&=&(a_{i}\mp1)^2-(a_{j}-a_{k})^{2},\nonumber\\
a_{\mp+}&=&(a_{i}\mp1)^2-(a_{j}+a_{k})^{2},
\end{eqnarray}
and $|a_{j}-a_{k}|+1\leq a_{i}\leq a_{j}+a_{k}-1$, where
$\{i,j,k\}$ run over all possible permutations of $\{1,2,3\}$.
The GR2 entanglement of a reduced state $\rho_{A_{i}A_{j}}$ of
modes $A_{i}$ and $A_{j}$ with CM $\gamma_{A_{i}A_{j}}$ reads as
\cite{Adesso_12}
\begin{equation}\label{GR2}
E_{2}^{G}\left(\rho_{A_{i}A_{j}}\right)=\frac{1}{2}\ln g_{k},
\end{equation}
where \cite{Adesso_05}
\begin{equation}\label{gk}
g_{k}=\left\{\begin{array}{lll} 1, & \textrm{if} & a_{k}\geq\sqrt{a_{i}^{2}+a_{j}^{2}-1};\\
\frac{\zeta}{8a_{k}^{2}}, & \textrm{if} & \alpha_{k}<a_{k}<\sqrt{a_{i}^{2}+a_{j}^{2}-1};\\
\left(\frac{a_{i}^2-a_{j}^2}{a_{k}^2-1}\right)^{2}, & \textrm{if}
& a_{k}\leq\alpha_{k}.
\end{array}\right.
\end{equation}
Here,
\begin{widetext}
\begin{eqnarray}
\alpha_{k}&=&\sqrt{\frac{2(a_{i}^{2}+a_{j}^{2})+(a_{i}^{2}-a_{j}^{2})^{2}+|a_{i}^{2}-a_{j}^{2}|\sqrt{(a_{i}^{2}-a_{j}^{2})^{2}+8(a_{i}^{2}+a_{j}^{2})}}
{2(a_{i}^{2}+a_{j}^{2})}},\label{alpha}\\
\delta&=&(-1+a_{1}-a_{2}-a_{3})(1+a_{1}-a_{2}-a_{3})(-1+a_{1}+a_{2}-a_{3})(1+a_{1}+a_{2}-a_{3})\nonumber\\
&&\times(-1+a_{1}-a_{2}+a_{3})(1+a_{1}-a_{2}+a_{3})(-1+a_{1}+a_{2}+a_{3})(1+a_{1}+a_{2}+a_{3}),\label{delta}\\
\zeta&=&-1+2a_{1}^2+2a_{2}^2+2a_{3}^2+2a_{1}^{2}a_{2}^{2}+2a_{1}^{2}a_{3}^{2}+2a_{2}^{2}a_{3}^{2}-a_{1}^{4}-a_{2}^{4}-a_{3}^{4}-\sqrt{\delta}.
\label{zeta}
\end{eqnarray}
\end{widetext}

A state with CM (\ref{GLEMS}), where $a>b$, is a two-mode reduced
state of its own three-mode purification with CM (\ref{gammapi}),
where block $\gamma_{AB}$ is given in Eq.~(\ref{GLEMS}), and
blocks $\gamma_{ABE}$ and $\gamma_{E}$ are given by
\begin{eqnarray}\label{gammaABEgammaEGLEMS}
\gamma_{ABE}=\left(\begin{array}{c}
\sqrt{(a-b)(a+1)}\sigma_{z} \\
\sqrt{(a-b)(b-1)}\openone \\
\end{array}\right),\quad \gamma_{E}=(1+a-b)\openone.\nonumber\\
\end{eqnarray}
Comparison of the purification with CM (\ref{threemodepure})
reveals that $a_{1}=a$, $a_{2}=b$, $a_{3}=1+a-b$, where we have
performed the following identification of modes of the
purification, $A\equiv A_{1}, B\equiv A_{2}$ and $E\equiv A_{3}$.
The evaluation of GR2 entanglement for states
$\bar{\rho}_{AB}^{(3)}$, $E_{2}^{G}(\bar{\rho}_{AB}^{(3)})$, thus
requires us to calculate $g_{3}$ given in Eq.~(\ref{gk}) and for
this we need to identify for which of the considered states first,
second and third branch on the RHS of Eq.~(\ref{gk}) applies.
Therefore, we have to investigate each branch separately.

1. For our states inequality
$a_{3}\geq\sqrt{a_{1}^{2}+a_{2}^{2}-1}$ reads as
$1+a-b\geq\sqrt{a^2+b^2-1}$, which is further equivalent with
inequality $b\leq1$. Since $b$ is a symplectic eigenvalue of a
quantum state, it satisfies inequality $b\geq1$, and therefore
inequality $b\leq1$ can only be satisfied for $b=1$. For these
states $g_{3}=1$ and consequently
$E_{2}^{G}(\bar{\rho}_{AB}^{(3)})=0$ by Eq.~(\ref{GR2}). The
vanishing of GR2 entanglement is in accordance with the fact that
CM (\ref{GLEMS}) for $b=1$ is a direct sum
$\bar{\gamma}_{AB}^{(3)}=(a\openone)\oplus\openone$, which
describes a product state which is separable.

2. From previous discussion it follows, that for the remaining
states it holds that $b>1$ and therefore the second inequality
$a_{3}<\sqrt{a_{1}^{2}+a_{2}^{2}-1}$ of the second branch on the
RHS of Eq.~(\ref{gk}) is always fulfilled. Further, a rather
lengthy algebra reveals that the first inequality
$\alpha_{3}<a_{3}$ is equivalent with inequality
\begin{equation}\label{inequality}
-8(a^2+b^2)(a-b)^{2}(a+1)(b-1)>0,
\end{equation}
which is never satisfied and therefore the second branch does not
apply for considered set of states.

3. Since according to previous discussion it always holds that
$\alpha_{3}\geq a_{3}$, where equality sign occurs either for
symmetric states with $a=b$ or states with $b=1$, one finds that
for GLEMS with CM (\ref{GLEMS}), where $a>b$, the GR2 entanglement
is given by Eq.~(\ref{GR2}), where $g_{3}$ is given by the third
branch on the RHS of Eq.~(\ref{gk}). This yields explicitly GR2
entanglement in the present case in the form
\begin{equation}\label{GR2GLEMS}
E_{2}^{G}\left(\bar{\rho}_{AB}^{(3)}\right)=\ln\left(\frac{a+b}{a-b+2}\right).
\end{equation}
As for $b=1$ the latter formula gives
$E_{2}^{G}(\bar{\rho}_{AB}^{(3)})=0$ it also covers the case 1.
Comparison of Eq.~(\ref{GIEGLEMS}) with Eq.~(\ref{GR2GLEMS})
unveils immediately that on the set of states with CM
(\ref{GLEMS}), where $a>b$ and $\sqrt{ab}\leq 2.41$, GIE is equal
to GR2 entanglement.

It remains to calculate GR2 entanglement for states
$\tilde{\rho}_{AB}^{(3)}$ with CM (\ref{tildeGLEMS}), where $a<b$.
The states possess three-mode purification with CM
(\ref{gammapi}), where block $\gamma_{AB}$ is given by the RHS of
Eq.~(\ref{tildeGLEMS}), whereas blocks $\gamma_{ABE}$ and
$\gamma_{E}$ are of the following form:
\begin{eqnarray}\label{gammaABEgammaEtildeGLEMS}
\gamma_{ABE}=\left(\begin{array}{c}
\sqrt{(b-a)(a-1)}\openone \\
\sqrt{(b-a)(b+1)}\sigma_{z} \\
\end{array}\right),\quad \gamma_{E}=(1+b-a)\openone.\nonumber\\
\end{eqnarray}
Consequently, the CM of the purification attains the form (\ref{threemodepure})
with $a_{1}=a$, $a_{2}=b$, $a_{3}=1+b-a$.

The discussion of the value of the parameter $g_{3}$,
Eq.~(\ref{gk}), for the case $a<b$ is exactly the same as in the
case $a>b$. The only differences are that in all discussed
inequalities as well as equalities the parameters $a$ and $b$ are
exchanged, and also that CM (\ref{tildeGLEMS}) for $a=1$ attains
the form $\tilde{\gamma}_{AB}^{(3)}=\openone\oplus(b\openone)$.
Hence, one finds that for states with CM (\ref{tildeGLEMS}), where
$a<b$, GR2 entanglement is given by formula (\ref{GR2GLEMS}),
where parameters $a$ and $b$ are exchanged, i.e.,
\begin{equation}\label{tildeGR2GLEMS}
E_{2}^{G}\left(\tilde{\rho}_{AB}^{(3)}\right)=\ln\left(\frac{a+b}{b-a+2}\right).
\end{equation}

Putting all the obtained results together we see, that for asymmetric
GLEMS with CM (\ref{gammast3}), where parameter $k$ is given in Eq.(\ref{k}),
the GR2 entanglement is equal to
\begin{equation}\label{GR2asymGLEMS}
E_{2}^{G}\left(\rho_{AB}^{(3)}\right)=\ln\left(\frac{a+b}{|a-b|+2}\right).
\end{equation}
Comparison of the latter equation with Eq.~(\ref{GIEGLEMSfinal})
again reveals that also for asymmetric states $\rho_{AB}^{(3)}$
satisfying condition $\sqrt{ab}\leq 2.41$, GIE is equal to the GR2
entanglement.

\begin{thebibliography}{99}
\bibitem{PES} In public Eve scenario \cite{Christandl_04}, Eve
must choose a function, which is applied to $E$ to obtain
$\tilde{E}$. The value of $\tilde{E}$ together with the
description of the function is then broadcasted to Alice and Bob.
\bibitem{CCM} A CCM is a real, symmetric and postive-semidefinite
matrix.
\end{thebibliography}
\end{document}